\documentclass[sigconf]{acmart}

\AtBeginDocument{%
  \providecommand\BibTeX{{%
    \normalfont B\kern-0.5em{\scshape i\kern-0.25em b}\kern-0.8em\TeX}}}

\acmConference[SIGMOD '21]{SIGMOD '21: ACM SIGMOD  International Conference on Management of Data}{June 20--25, 2018}{Xi'an, Shaanxi, China}
\acmBooktitle{SIGMOD '21: ACM SIGMOD International Conference on Management of Data,
  June 20--25, 2021, Xi'an, Shaanxi, China}
\acmPrice{15.00}
\acmISBN{978-1-4503-XXXX-X/18/06}

\usepackage{xspace}
\usepackage{wrapfig}
\newcommand{\red}[1]{\textcolor{red}{#1}}
\newcommand{\blue}[1]{\textcolor{blue}{#1}}
\newcommand{\projectname}{RCC\xspace}
\newcommand{\nowait}{\textsf{\small{NOWAIT}}\xspace}
\newcommand{\waitdie}{\textsf{\small{WAITDIE}}\xspace}
\newcommand{\occ}{\textsf{\small{OCC}}\xspace}
\newcommand{\mvcc}{\textsf{\small{MVCC}}\xspace}
\newcommand{\sundial}{\textsf{\small{SUNDIAL}}\xspace}
\newcommand{\calvin}{\textsf{\small{CALVIN}}\space}
\def\step{step } % used for defining every step in cc_algs, can change
\begin{document}

%%
%% The "title" command has an optional parameter,
%% allowing the author to define a "short title" to be used in page headers.
\title{Comprehensive Framework of RDMA-enabled Concurrency Control Protocols}

%%
%% The "author" command and its associated commands are used to define
%% the authors and their affiliations.
%% Of note is the shared affiliation of the first two authors, and the
%% "authornote" and "authornotemark" commands
%% used to denote shared contribution to the research.
% \author{Ben Trovato}
% \authornote{Both authors contributed equally to this research.}
% \email{trovato@corporation.com}
% \orcid{1234-5678-9012}
% \author{G.K.M. Tobin}
% \authornotemark[1]
% \email{webmaster@marysville-ohio.com}
% \affiliation{%
%   \institution{Institute for Clarity in Documentation}
%   \streetaddress{P.O. Box 1212}
%   \city{Dublin}
%   \state{Ohio}
%   \postcode{43017-6221}
% }

\author{Chao Wang}
\affiliation{%
  \institution{University of Southern California}
  %\streetaddress{1 Th{\o}rv{\"a}ld Circle}
  %\city{Hekla}
  %\country{Iceland}
  }
\email{wang484@usc.edu}

\author{Kezhao Huang}
\affiliation{%
  \institution{Tsinghua University}
  %\city{Rocquencourt}
  %\country{France}
}
\email{kz16@mails.tsinghua.edu.cn}

\author{Xuehai Qian}
\affiliation{%
  \institution{University of Southern California}
  %\city{Rocquencourt}
  %\country{France}
}
\email{xuehai.qian@usc.edu}

\begin{abstract}

In this paper, we develop
\projectname, the first
unified and comprehensive RDMA-enabled distributed transaction processing framework supporting six serializable concurrency control protocols---not only the classical protocols \nowait, \waitdie and \occ, but also more advanced \mvcc and \sundial, and 
even \calvin---the deterministic concurrency control protocol.
Our goal is to unbiasedly compare the protocols 
in a common execution environment with the concurrency 
control protocol being the only changeable component. 
We focus on the correct and efficient implementation
using key techniques, i.e., 
co-routines, outstanding requests, and
doorbell batching, with two-sided and one-sided 
communication primitives. 
Based on \projectname, we get the deep insights
that cannot be obtained by any existing systems.
Most importantly, we obtain the execution stage latency 
breakdowns with one-sided and two-sided primitive 
for each protocol, which are
analyzed to develop more efficient 
hybrid implementations. \projectname also supports the enumeration of all stage-wise hybrid designs under given workload characteristic.
Our results show that three hybrid designs are indeed better than both one-sided and two-sided implementations by up to 17.8\%.
We believe that \projectname is a significant advance
over the state-of-the-art; it can both provide performance insights
and be used as the common infrastructure for fast
prototyping new implementations. 

\end{abstract}
%%
%% The code below is generated by the tool at http://dl.acm.org/ccs.cfm.
%% Please copy and paste the code instead of the example below.
%%

% \ccsdesc[500]{Computer systems organization~Embedded systems}
% \ccsdesc[300]{Computer systems organization~Redundancy}
% \ccsdesc{Computer systems organization~Robotics}
% \ccsdesc[100]{Networks~Network reliability}

%%
%% Keywords. The author(s) should pick words that accurately describe
%% the work being presented. Separate the keywords with commas.

%% A "teaser" image appears between the author and affiliation
%% information and the body of the document, and typically spans the
%% page.
% \begin{teaserfigure}
%   \includegraphics[width=\textwidth]{sampleteaser}
%   \caption{Seattle Mariners at Spring Training, 2010.}
%   \Description{Enjoying the baseball game from the third-base
%   seats. Ichiro Suzuki preparing to bat.}
%   \label{fig:teaser}
% \end{teaserfigure}

%%
%% This command processes the author and affiliation and title
%% information and builds the first part of the formatted document.
\maketitle

% \input{abstract}
%\vspace{-2mm}
\section{Introduction}
%\vspace{-1mm}
%\vspace{-1mm}
%-- The importance and challenges of distributed 
% transactions. (from VLDB 1st and 2nd paragraph of introduction)

On-line transaction
processing (OLTP) has ubiquitous 
applications in many important domains, including
banking, stock marketing, e-commerce, etc. 
As the data volume grows exponentially, 
single-server systems
experience major difficulties in handling a large number of queries from clients due to limited system resources. Thus, partitioning data sets across distributed machines is
necessary and becoming increasingly important. 
%becoming increasingly necessary and important. 
However, partitioning data such that all queries access only one partition is challenging~\cite{curino2010schism,pavlo2012skew}. 
In practice, transactions inevitably access multiple networked machines.

Distributed transactions should guarantee
two key properties:
(a) atomicity: either all or none of the machines agree to apply the updates; and (2) serializability: all transactions must commit in some serializable order. 
%must have the following properties 1) atomicity. Either all machines agrees to apply the changes or none of them does. 2) serializability. All transactions must commit changes as if they commit in some serializable order. 
To ensure these properties, concurrency control 
protocols have been investigated for 
decades~\cite{agrawal1987concurrency,Bernstein:1981:CCD:356842.356846,bernstein1983multiversion,kung1981optimistic,mahmoud2014maat,tu2013speedy}.
%there has been a large amount of studies in the database community about distributed transactions since the 1980's. 
%Prior works have proposed many distributed concurrency control protocols such as two-phase Locking (2PL)~\cite{Bernstein:1981:CCD:356842.356846}, time\-stamp-based\-~\cite{Bernstein:1981:CCD:356842.356846}, multi-versioned\-~\cite{bernstein1983multiversion}, opti\-mistic protocols (OCC)\-~\cite{kung1981optimistic}, etc. % MaaT~\cite{mahmoud2014maat}etc. 
The well-known challenge of multi-partition serializable concurrency control protocols
is the significant performance 
penalty due to the communication and coordination among distributed machines~\cite{lee2015implementing,stonebraker1986case,Thomson:2012:CFD:2213836.2213838}.
When a transaction accesses multiple records over the network,
it needs to be serialized with all conflicting 
transactions~\cite{bailis2014coordination}.
%any other transactions it conflicts with have to be serialized~\cite{bailis2014coordination}. 
Therefore, a high-performance network is crucial.
%for distributed transactions.

% -- Performance advantage of RDMA and fast network. (1st paragraph of OSDI in introduction) 

Remote Direct Memory Access (RDMA) is a
new technology that enables the network interface card (NIC) to access the memory of a 
remote servers in a distributed cluster. 
Due to its high bandwidth and low
latency, RDMA has been recently used to support 
distributed transaction systems~\cite{10.1145/2619239.2626299,wei2015fast,kalia2016fasst,chen2016fast,dragojevic2015no} and enhanced the performance by orders of magnitude compared to traditional systems using TCP. 
RDMA network supports both TCP-like 
{\em two-sided} communication using primitives \texttt{SEND/RECV}, and
{\em one-sided} communication using 
primitives \texttt{READ/WRITE/ATOMIC},
which are capable of accessing remote memory while bypassing traditional network stack,
the kernel, and even the remote CPUs. 
% To understand the performance of these primitives, there have already been 
% intensive studies investigating the 
% pros and cons of using each primitive. 

Extensive studies have been conducted in understanding
the performance implication of each primitive using micro-benchmarks~\cite{10.1145/2619239.2626299,dragojevic2014farm,dragojevic2015no,wei2018deconstructing,tsai2017lite}. 
%the pros and cons of using each primitive to understand their performance implications.
Moreover, RDMA has been used to implement the 
Optimal Concurrency Control (\occ) protocol~\cite{kalia2016fasst,dragojevic2014farm,wei2018deconstructing}.
Two takeaways from DrTM+H~\cite{wei2018deconstructing} are: 
(1) the best performance of \occ cannot be simply achieved by solely using two-sided or one-sided communication; and 
(2) different communication primitives 
are best suited for each execution stage. 
%suitable for different protocol {\em stages}. 
They suggest 
that achieving the optimal performance 
of a concurrency control protocol using RDMA is far from trivial and calls for a systematic investigation. 
Second, building the standalone framework for each individual protocol does not allow fair and unbiased cross-protocol
comparison 

We claim that the state-of-the-art RDMA-based system DrTM+H~\cite{wei2018deconstructing} is {\em not} sufficient
for two important reasons. 
First, in real-world applications, various concurrency
control protocols~\cite{mu2014extracting,mahmoud2014maat,roy2015homeostasis,escriva2015warp,cowling2012granola,corbett2013spanner,faleiro2014rethinking} are used,
the understanding of RDMA implications on \occ
does {\em not transfer} to other protocols.
Second, building the standalone framework for 
each individual protocol does not allow the fair and unbiased
{\em cross-protocol} performance comparison.
In a complete system for 
execution distributed transactions, concurrency control
protocol is only one component, the system organization, optimizations, and transaction execution model
can vary a lot.
Having a common execution 
environment for all various protocols is critical to draw any
meaningful conclusions~\cite{yu2015evaluation,yu2014staring}. 
Unfortunately, DrTM+H does not provide such capability.
Compared to DrTM+H, 
Deneva~\cite{harding2017evaluation} studied
six concurrency control protocols based on TCP, 
affirming the importance of cross-protocol 
comparison. However, Deneva is not based on RDMA.

In this paper, we take the important step to close
the gap. We develop {\em \projectname}, the {\em first}
unified and comprehensive RDMA-enabled distributed transaction processing framework supporting multiple 
concurrency control protocols with different properties. 
Currently, it includes protocols in a wide spectrum:
(1) classical protocols such as two-phase-locking (2PL), i.e.,
\nowait~\cite{Bernstein:1981:CCD:356842.356846} and
\waitdie~\cite{Bernstein:1981:CCD:356842.356846}, and 
\occ~\cite{kung1981optimistic}, of which RDMA-based 
implementations have been studies thoroughly;
(2) more advanced protocols such as \mvcc~\cite{bernstein1983multiversion}, which has been adopted by modern high-performance database systems, and 
the recent \sundial~\cite{yu2018sundial}, that allows
dynamically adjustment of commit order with logical lease
to reduce abort; and
(3) the deterministic protocol \calvin~\cite{Thomson:2012:CFD:2213836.2213838}, a shared-nothing protocol that ensures deterministic transaction execution.

\projectname enables us to perform 
{\em unbiased and fair} comparison of the protocols 
in a {\em common} execution environment with the concurrency 
control protocol being the only changeable component. 
%While the protocol specifications are known, 
We develop the {\em correct and efficient} RDMA-based
implementation using known techniques, i.e., 
co-routines, outstanding requests, and
doorbell batching, with two-sided and one-sided 
communication primitives. 
To validate the benefits of RDMA, \projectname also provides
reference implementations based on TCP.
As a common infrastructure
for RDMA-enabled distributed transaction execution, 
\projectname allows the fast prototyping of
other existing protocols or {\em new implementations}.

We believe \projectname is a significant advance
of the state-of-the-art for three reasons. 
First, while the protocol specifications are known, 
we answer the question of {\em how} to leverage
RDMA to construct different protocols with 
concrete, executable, and efficient implementations. 
Second, we can perform both
apple-to-apple {\em cross-protocol} comparisons and 
the {\em stage-level same-protocol} study
on performance and various execution characteristics
in the context of the same system organization. 
The observations of which primitives being best
suited for which execution stage
can be used to further optimize the
performance.
Third, for \calvin, which is a shared-nothing protocol 
and has never been studied in the context of RDMA, 
we answer the question of whether 
the one-sided primitives would bring the similar benefits
as other shared-everything protocols.
In summary, with \projectname we can get the deep 
insights that cannot be obtained by any
existing systems.

The implementation of the current \projectname with the 
six protocols has around 25,000 lines of codes written in C++.
We intend to open-source the framework in the near future. 
We try our best to fairly optimize the performance
of each without bias
using known techniques such as co-routines~\cite{kalia2016fasst}, outstanding requests~\cite{wei2018deconstructing},
doorbell batching~\cite{kalia2016design}.
We evaluate all protocol implementations on a cluster with ConnectX-4 EDR InfiniBand RDMA support using three typical workloads:
%We evaluate all protocols in \projectname on {\em two clusters} with different  RDMA network capabilities.One is equipped with ConnectX-4 EDR 100Gb/s InfiniBand (EDR) and the other with ConnectX-3 Pro FDR 56Gb/s InfiniBand (FDR). We use them due to the huge difference in supporting RDMA one-sided operations: the latency of a single one-sided operation using FDR InfiniBand can be 5x more than that of using the EDR InfiniBand. We evaluate \projectname on three OLTP workloads: 
SmallBank~\cite{SmallBank},  TPC-C~\cite{TPC-C}, and
YCSB~\cite{cooper2010benchmarking}

We perform the {\em first cross-protocol} performance comparison with RDMA and
observe that \occ does not always achieve the best performance. 
In fact, the simple 2PL protocols such as \nowait and \waitdie perform well with high performance RDMA. 
Most importantly, we obtain the execution stage latency 
breakdowns with one-sided and two-sided primitive 
for each protocol for all three workloads. They can be analyzed to develop {\em hybrid} implementations, which 
may achieve better performance under the given a workload characteristic. 
Our experiment shows that by cherry-picking the communication type that incurs lower latency for each protocol stage, we can find new protocol implementations that reaches 
at most 17.8\% speedup, compared to the better 
implementation using RPC or one-sided primitives.

%We present the analysis methodology to determine the 
%best primitives to use for each stage. 
With a simple interface, \projectname allows both common and advanced users to quickly evaluate any hybrid implementation for an existing or new protocol given a workload characteristic. In addition, for a given protocol, \projectname can exhaustively enumerate all combinations of hybrid protocols and provide substantial evidence that a certain hybrid design is the best among all possibilities when varying stage communication styles.
We believe that \projectname is a significant advance
over state-of-the-art, it can both provide performance insights
and be used as the common infrastructure for fast
prototyping new implementations.

\section{Background}
%

%\vspace{-2mm}
%\subsection{RDMA and Its Primitives}
% \vspace{-2mm}

%-- Similar to OSDI's corresponding section. 

Remote Direct Memory Access (RDMA) is a network technology featuring high bandwidth and 
low latency data transfer with low CPU overhead. 
%transfer data in  low latency and high bandwidth while incurring low CPU overhead. 
It is widely considered suitable for large data centers. RDMA operations, i.e., \textbf{verbs}, can be classified into two types:
(1) {\em two-sided} primitives 
\texttt{SEND/RECV}; and 
(2) {\em one-sided} primitives 
\texttt{READ/WRITE/ATOMIC}.
The latter provides the unique capability to directly access the memory of remote machines without involving remote CPUs. 
This feature makes one-sided operations suitable for distributed applications with high CPU utilization. 
Although having similar semantics with TCP's send/receive over bound sockets,
%These are the build in .. for tpc (\blue{TODO: there is a sentence in another section which describe the relationship between tcp and two-sided well, i didn't find it}). 
RDMA two-sided operations bypass the traditional network stack and the OS kernel, 
making the performance of 
RPC implementation over RDMA much higher than
that over TCP.

%The first type is one-sided operations including RDMA READ, RDMA WRITE and RDMA ATOMIC. One-sided operations are special in RDMA since they can directly access the memory of a remote machine without involving the remote CPU. This feature makes one-sided operations suitable for distributed applications with high CPU utilization. 
%The other type is two-sided RDMA operations, including RDMA SEND and RDMA RECV. Although having similar semantics with TCP's send/receive over bound sockets,
%These are the build in .. for tpc (\blue{TODO: there is a sentence in another section which describe the relationship between tcp and two-sided well, i didn't find it}). 
%they bypass the tranditional network stack and the OS kernel, making the RPC implementation over RDMA outperform that over TCP/IP.

To perform RDMA communication, 
queue pairs (QPs) must be set up. 
A QP consists of a send queue and a receive queue. When a sender posts a one-sided RDMA request to the send queue, the local QP will transfer data to some remote QP, and the sender can poll for completion information from the completion queue associated with the QP.
%once the operation is completed. 
The receiver's CPU is not aware of the one-sided operations performed by the receiver's RNIC without checking the changes in memory. For a sender to post a two-sided operation, the receiver QP has to post \texttt{RECV} for the corresponding \texttt{SEND} in advance. It polls the receive queue to obtain the data.
%QP has three different transport modes: Reliable Connection (RC), Unreliable Connection (UC), and Unreliable Datagram (UD). One RC QP can send messages reliably to exactly one another connected RC QP; a UD QP can either unicast to one or multicast to many UD QPs without delivery guarantee. One-sided operations are only fully supported between two RC QPs, while two-sided operations must be used for UD QPs. 
To set up a reliable connection, a node has to maintain at least a cluster-size number of QPs in its RDMA-enabled NIC (RNIC), each connected with one remote node. 
%On the contrary, a node needs to maintain only one UD QP to send/receive data across the cluster, thus saving system resources.

%\vspace{-2mm}
%\subsection{RDMA-enabled Distributed Transactions}
% \vspace{-2mm}
%-- Similar to OSDI's corresponding section. 

%Distributed transaction systems are popular applications with demanding network operations. 
Prior works studied 
employing RDMA for distributed transactions. \cite{chen2016fast} uses only one-sided operations to transfer and update records. \cite{kalia2016fasst} uses UD to implement RPC in its transaction framework. \cite{wei2018deconstructing} 
proposes a hybrid implementation that 
uses one-sided and two-sided operations for different stages of transactions. All these frameworks focus on \occ~\cite{kung1981optimistic,huang1991experimental}.

\section{\projectname System Organization}
%\vspace{-2mm}
%-- Similar to Section 2 of VLDB, need a figure similar to Figure 1 of that paper

%\vspace{-2mm}
\subsection{Overall Architecture}
% \vspace{-2mm}

%-- Based on the figure, explain the overall design flow shared by all protocols. I understand that the design is based on ROCC, so the figure will be  different from VLDB. You need to draw a figure based on ROCC framework and describe the common interface and components for all protocols. 

%In order to implement and evaluate representative concurrency control protocols using RDMA primitives, we designed a framework called \textbf{\system}. 
\setlength{\intextsep}{2pt}%
\setlength{\columnsep}{8pt}%
% \begin{wrapfigure}[10]{r}{0.55\linewidth}
\begin{figure}[t]
    \centering
    % \vspace{-0.2cm}
    \includegraphics[width=0.8\linewidth]{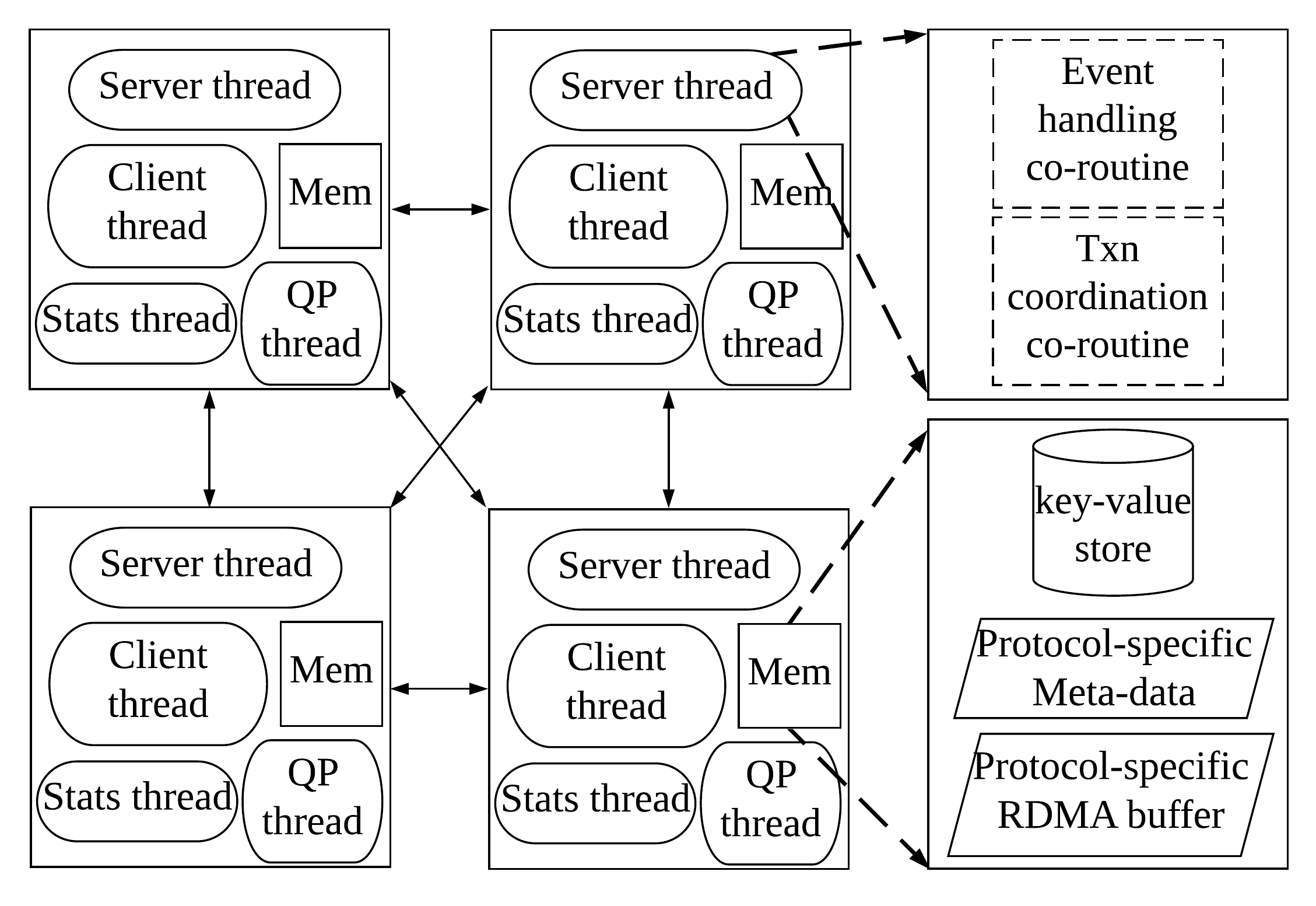}
    \vspace{-2mm}
    \caption{RCC Framework Overview}
    \vspace{-0.2cm}
    \label{fig:system}
% \end{wrapfigure}
\end{figure}

Figure~\ref{fig:system} shows the overview of 
\projectname, which runs on multiple symmetric distributed nodes,
each containing a configurable number of server threads 
to process transactions. A client thread sends transaction requests to a random local or remote transaction processing thread in the cluster. The stats thread is used to collect the statistics (e.g., the number of committed transactions) generated by each processing thread. The QP thread is used to bootstrap RDMA connections by establishing the pairing of RDMA QPs using TCP connections. 

\projectname uses co-routines as an essential optimization technique~\cite{kalia2016fasst} to hide network latency. 
Specifically, each thread starts an {\em event handling} co-routine and some {\em transaction coordination} co-routines. An event handling co-routine continuously checks and handles network-triggered events such as polled completions or {memory-triggered events} such as the release of a lock. A transaction coordination co-routine is where a transaction logically executes.

In \projectname, the distributed in-memory database is implemented as a distributed key-value store that can be accessed either locally or remotely via a key and table ID. we leveraged DrTM+H's\cite{wei2018deconstructing} key-value store as \projectname's back-end. In addition to the in-memory database, each protocol has its protocol-specific metadata or RDMA buffer to ensure the correct execution when leveraging RDMA primitives.

%\vspace{-2mm}
\subsection{Transaction Execution Model}
% \vspace{-2mm}

%-- See VLDB paper, but our model is **not** based on stored procedure. Explain our model here: how data are distributed? server/client. (Also check OSDI paper, it also mentioned ``symmetric model''..)
\projectname employs a symmetric model to execute transactions: each node serves as both a client and a transaction processing server. As shown in Figure~\ref{fig:system}, each transaction coordination co-routine is responsible for executing a transaction at any time. 
We use {\em coordinator} to refer to the co-routine that receives transaction requests from some local or remote client thread and orchestrates transactional activities in the cluster. We use {\em participant} to refer to a machine where there is a record to be accessed by some transaction. When a 
participant receives an RPC request, its event handling co-routine will be invoked to process the request locally.
When a participant receives an RDMA one-sided 
operation, its RNIC is responsible for accessing the memory without interrupting the CPU.

\begin{figure}[t]
    \centering
    \vspace{-2.5mm}
    \includegraphics[width=\linewidth]{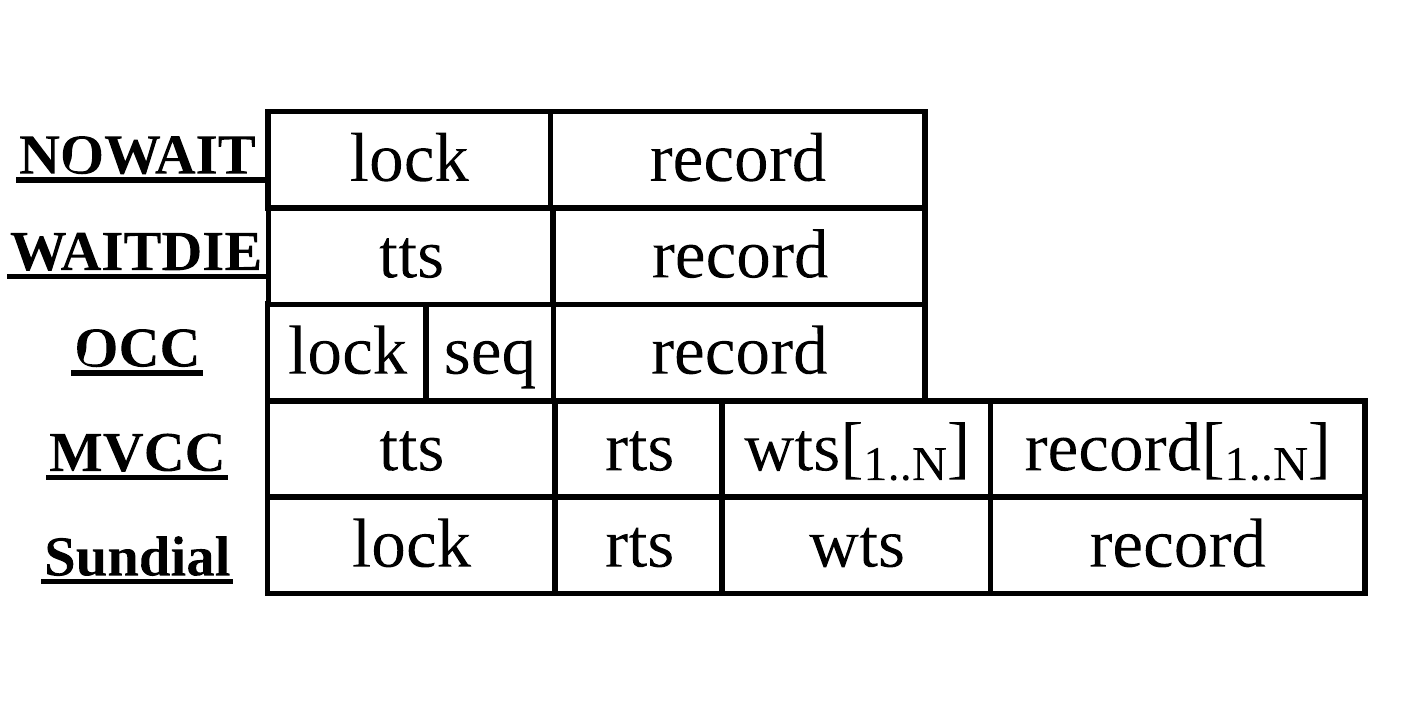}
    \vspace{-8mm}
    \caption{Protocol Metadata}
    \vspace{-4mm}
    \label{fig:metadata}
% \end{wrapfigure}
\end{figure}

In \projectname, A {\em record} refers to the
actual data; and a {\em tuple} refers to a record associated with the relevant metadata. All tuples are located in RDMA-registered memory. A distributed in-memory key-value store keeps all tuples partitioned among all machines. Since one-sided operations can only access remote memory by leveraging the pre-computed remote offsets, to reduce the number of one-sided operations involved in retrieving metadata, the metadata are placed physically together with the record as shown in Figure~\ref{fig:metadata}. Currently, \projectname only supports fixed record size and variable-sized record can be supported by placing an extra pointer in the record field pointing to an RDMA-registered region, similar to\cite{zamanian2016end}. With one-sided \texttt{READ}, the remote offset of a tuple is fetched before the actual tuple is fetched and the offset is then cached locally to avoid unnecessary one-sided operations.

A transaction has a {\em read set (RS)} and a {\em write set (WS)}
that are known before the execution. 
The records in RS are read-only.
%contains reads and writes of 
%records that need to be performed atomically.
The execution of a transaction is
conceptually divided into three primary stages: 
% \begin{comment}
% ~\footnote{We use this assumption for simplicity. The implementation
% can also be applied to transactions with multiple fetching and execution stages.}
% \end{comment}
1) {\em fetching}: get the tuples of records in RS and WS, the metadata
is used for protocol operations;
%where a read/write operation fetches the record and/or metadata if needed;
2) {\em execution}: a transaction performs the actual computation locally using the fetched record; and
3) {\em commit}: a transaction checks if it is serializable, if so, {\em logs} all writes to remote backup machines for high availability and recovery, and {\em updates} remote records.
Our implementations can be applied to transactions with one or more fetching and execution stages.

\begin{comment}
In our \projectname, 
%we assume that each transaction will only have one commit stage, which means the transaction will end after committing. To make our introduction more concise, 
we assume that there is just one data fetching and execution stage. However the implementation
can be applied to transactions with multiple date fetching and execution stages.
The execution of each read and write is
composed of three stages:
1) {\em access} the data and add the operation into {\em read or write set};
2) {\em validate} the operation; and 
3) {\em commit} the write set when the transaction

Generally, a transaction execution is divided into three stages, and consists of write or read operations. The first stage is \textbf{fetching}, where the write or read operation  fetches the needed record. The second stage is \textbf{execution}, where the transaction will do the actual computation locally using the record just fetched. The third stage is \textbf{commit}. In this stage, the transaction checks the metadata to find out if the execution was valid. after which it commits the updated records as well as their corresponding metadata.
\end{comment}

%\vspace{-2mm}
\subsection{RDMA Communication and Optimizations}
% \vspace{-2mm}

% -- See 3.1 and 3.2 of OSDI, try to be brief since they are not our contribution. Also mention any optimizations we have proposed and validated to work well. 

% \setlength{\intextsep}{2pt}%
% \setlength{\columnsep}{8pt}%
%  \begin{wrapfigure}[8]{r}{.65\linewidth}
\begin{figure}[htp]
    \centering
    \vspace{-0.2cm}
    \includegraphics[width=\linewidth]{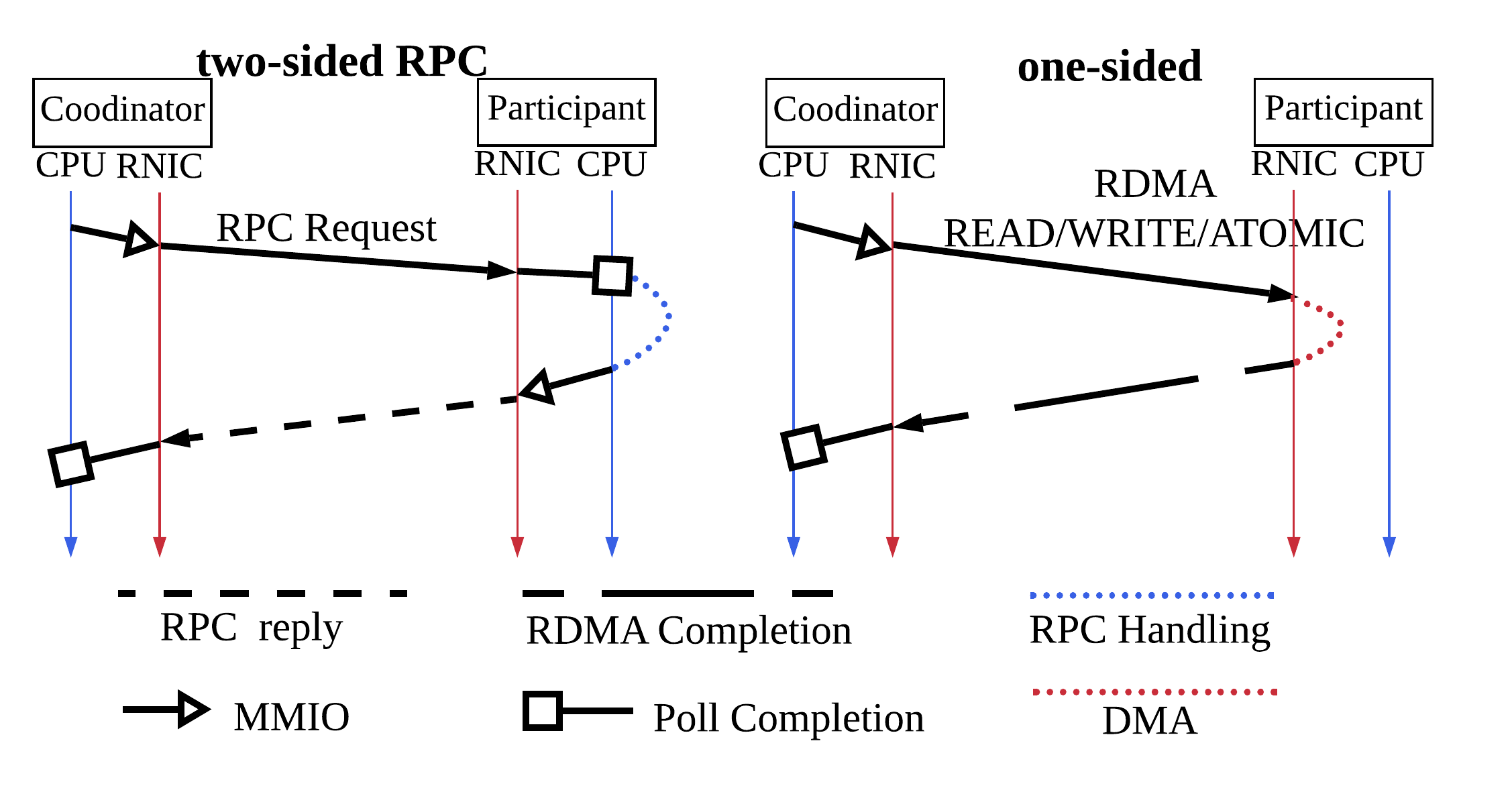}
    \vspace{-0.6cm}
    \caption{Two-sided versus one-sided}
    \vspace{0.1cm}
    \label{fig:two-vs-one-primitives}
%  \end{wrapfigure}
\end{figure}
We use two-sided RDMA primitives over UD QPs to implement RPC. From ~\cite{kalia2016fasst},
two-sided primitives over UD QPs outperform one-sided primitives in symmetric transaction systems, and UD mode is much more reliable than expected with RDMA network's lossless link layer. \cite{wei2018deconstructing} further confirms the unsuitability of one-sided primitives to implement fast RPC. %compared to two-sided ones. 

%  \setlength{\intextsep}{2pt}%
%  \setlength{\columnsep}{8pt}%
%  \begin{wrapfigure}[5]{r}{0.45\linewidth}

Figure~\ref{fig:two-vs-one-primitives} illustrates the two types of communications in \projectname employed by each concurrency control protocol. In two-sided RPC, a coordinator first sends a memory-mapped IO (MMIO) to the RNIC, which in turn \texttt{SEND}s an RPC request to the receiver's RNIC. After the corresponding participant \texttt{RECV}s the request, its CPU polls a completion event, which later triggers a pre-registered handler function to process the request and send back a reply using similar verbs. In one-sided communication, after the participant receives a one-sided request, i.e., \texttt{READ}, \texttt{WRITE}, \texttt{ATOMIC}, its RNIC will access local memory using a Direct Memory Access (DMA). The completion is signaled when the coordinator polls if it is interested in the completion event.

MMIO is an expensive operation to notify RNIC of a request fetching event. Using {\em one} MMIO for a batch of RDMA requests can effectively save PCIe bandwidth and improve the performance of transaction systems~\cite{wei2018deconstructing}, which is called \emph{doorbell batching}. Meanwhile, having multiple outstanding requests on the fly can  save the waiting time of request completion, thus reducing the latency of remote transactions~\cite{wei2018deconstructing}. Leveraging co-routines serve to interleave RDMA communication with computation. \projectname uses similar techniques as important optimizations.
\section{RDMA-Based Concurrency Control}
% \vspace{-2mm}

%-- Describe each protocol design in detail, explain the ideas well. If you made some design choices based some considerations or tradeoffs, please also put them here. 

%-- Do not mention OCC as the baseline, just discuss
%all protocols in some order. When talking about OCC, mention that it is based on the OSDI (with any modifications). 

%-- For each protocol, remember to describe both: RPC and (mostly) RDMA. Draw some figures if needed, especially for the protocols that incurred significant coding/debugging (e.g. Calvin). 

%-- I remember that you mention certain operations are needed by all (or several) protocols and it is best to use RPC/RDMA to implement them. Please discuss them in the final "Discussion" section, and perhaps add a table with {operations, protocols} indicating any design options and findings in each cell. 

In \projectname, we implement six concurrency control protocols with two-sided and one-sided RDMA primitives. 
%The implementations of these protocols
%involve a variety of techniques. 
%The concurrency control algorithms that we primarily concern and included in our system cover a diversity of different techniques. 
%These protocols can be classified into several types. 
Among these protocols,
\nowait~\cite{Bernstein:1981:CCD:356842.356846} and \waitdie~\cite{Bernstein:1981:CCD:356842.356846} are
examples of two-phase locking (2PL)~\cite{Bernstein:1981:CCD:356842.356846} 
concurrency control algorithms.
They differ in conflict resolution,i.e.,
how conflicts are resolved to ensure serialization. 
%We will have a detailed discussion in their respective subsections. 
Compared to 2PL, Optimistic Concurrency Control (\occ)~\cite{kung1981optimistic} reads records speculatively without locking and 
validates data upon transaction commits---the only time to use locks. 
%OCC constraints locking only to the time when transactions commit. 
\mvcc~\cite{bernstein1983multiversion} optimizes the performance of read-heavy transactions by allowing the read of 
the correct recently committed records instead of aborting. \sundial~\cite{yu2018sundial} 
leverages the dynamically adjustable logical leases to 
order transaction commits and reduce aborts. 
\calvin~\cite{Thomson:2012:CFD:2213836.2213838} introduces determinism 
with a shared-nothing protocol, which demonstrates very different 
communication behavior. 
%\red{WRITE A SENTENCE ABOUT THE ADVANTAGE OF CALVIN}.

While the protocols themselves are known, 
we rethink their correct and efficient implementations
in the context of RDMA.
%\projectname is to rethink their {\em implementations and performance in the context of RDMA}. 
Each protocol
requires techniques to implement specific 
protocol requirements, particularly atomic tuple read (for \mvcc) and 
update (for \sundial). 
In this section describe two implementations of each protocol:
%For each protocol, we implement two versions:
1) RPC version, which mostly uses remote function call enabled by RDMA's two-sided communication primitives; 
and 2) one-sided version, which mainly uses RDMA's unique 
one-sided communication primitives. 
%Note that each version does not solely use
%one type of RDMA primitives---we choose the proper primitives
%to implement certain operations when the alternative
%is overwhelmingly worse. 
We will propose a hybrid design based on the stage latency
results of each protocol generated by \projectname.
%Next, we first discuss the common
%transaction structure and operations. 

%Currently, there are two major means of inter-nodes communication using RDMA. 1) Remote Procedure Call (RPC) using two-sided RDMA primitives (i.e.,  SEND/RECV). This is a natural drop-in replacement of RPC using TCP. 2) One-sided primitives such as WRITE/READ/ATOMIC operations. Figure~\ref{fig:two-vs-one-primitives} illustrates these two different types of communications.

%\vspace{-2mm}
\subsection{Transaction Operations}
% \vspace{-2mm}
% TODO: the name of stages in one operation and in one transaction
 
%is committed.
%In the implementation of each protocol by RDMA
%primitives, we should consider the trade-offs
%between one- and two-sided communication. 
\begin{comment}
The trade-off between two-sided and one-sided 
RDMA primitives are well-known. 
In the RPC version, 
operations are sent to and executed on the
remote machine. 
The advantage is that the whole operation can be done with one function call, thus saving communication.  
But the local machine needs to wait for the remote execution, which is typically done by
a co-routine after the remote CPU is interrupted.
In contrast, one-sided primitives 
bypass remote CPU and usually achieve
better performance for 
applications with higher CPU usage.
However, the local machine may launch more network requests and needs to maintain metadata for remote offsets. 
%whenever remote data is needed. 
%Moreover, metadata needs to be maintained to keep
%remote offsets. 
Also, one-sided atomic
operation tends to be costly. 
\end{comment}
We consider the following common operations 
used in one or multiple concurrency control protocols. 
They can be implemented with either RPC or one-sided
primitives.

{\bf Fetching.} Tuples are fetched during transaction execution. 
The read-only records are fetched into RS, other accessed records
are fetched into WS.
%If the transaction logic only intends to read the fetched record without modifying its value, the read operation fetches the value into a read set; otherwise, the write operation fetches the value into a write set.}

{\bf Locking.} All \projectname protocols need locking to enforce certain logical serialization order. For remote locking, 
the better implementation choice is 
affected by the load of remote threads which execute transaction co-routines. The higher load may affect the capability of handling RPC, 
thus one-sided primitives can be better.

%leads to weaker capability of RPC handling.

%If the remote machine has longer co-routine
%execution time, the remote lock handler in the RPC version will wait longer, and one-sided implementation is better. 
%We observe this behavior in YCSB 
%in which the co-routine
%execution time can be configured in the benchmark. 

%For TPC-C and SmallBank, RPC implementation is better
%since the co-routine execution time is short. 

{\bf Validation. } This operation is needed
in \occ, \mvcc, and \sundial in different stages.
The RPC implementation typically requires only one network operation,
while the one-sided version may lead to
one or more requests. Similar to locking, the best primitive choice is determined by the workload of remote co-routines.
%Similar to DrTM+H and FaSST, RCC employs the coordinator log\cite{10.1007/BF01264014} as the implementation of the two-phase-commit for simplicity: 

%but the RPC execution time needs to be considered.
%Similar to locking, the best choice is determined 
%by the co-routine execution time in remote machine.
%{\bf Two-phase-commit(2PC)} For all distributed concurrency control protocols, 2PC is a consensus protocol that ensures atomic commitment: it ensures that all participants agree either to commit or abort a distributed transaction. 2PC incurs non-negligible cost and needs to be enforced in all RPC and one-sided designs. 
%\projectname implements both the RPC and one-sided versions of 2PC for all concurrency control protocols. 
%\red{In comparison, DISCUSS ROCC DOES NOT HAVE THIS, AND THE CONSEQUENCE..}
%Since 2PC is independent of concurrency control protocols, we do not show it in protocol implementations in this section. Rather, 
%\red{We will discuss 2PC implementation shortly before 
%describing the details of each protocol.}

{\bf Logging} To support high availability and recovery, each protocol logs its updates to some backup servers. 
Similar to DrTM+H and FaSST, RCC employs coordinator log\cite{10.1007/BF01264014} for two-phase-commit. 
Only after the successful logging and reception of acknowledgments from all replica, can the transaction writes the updates back to the remote machine. Logs are lazily reclaimed in the background of backup machines when they are notified by the coordinator using two-sided RPC. Logging strongly prefers one-sided \texttt{WRITE} to log to backup servers for \occ according to \cite{wei2018deconstructing}. Our stage-wise latency
results support this claim for other protocols.

{\bf Update} It writes back the updated data and metadata. Two-sided RPCs can finish this update in one round trip; one-sided primitives need two without doorbell batching. The index of the write set entries can be cached in advance to reduce the overhead of this operation when using one-sided primitives.

% one-sided 
%implementation is better since the operation typically needs to write updated data in a remote machine. Moreover, the overhead of accessing metadata can be avoided by caching them in advance.

\begin{figure}[htp]
    \centering
    % \vspace{-0.2cm}
    \includegraphics[width=0.8\linewidth]{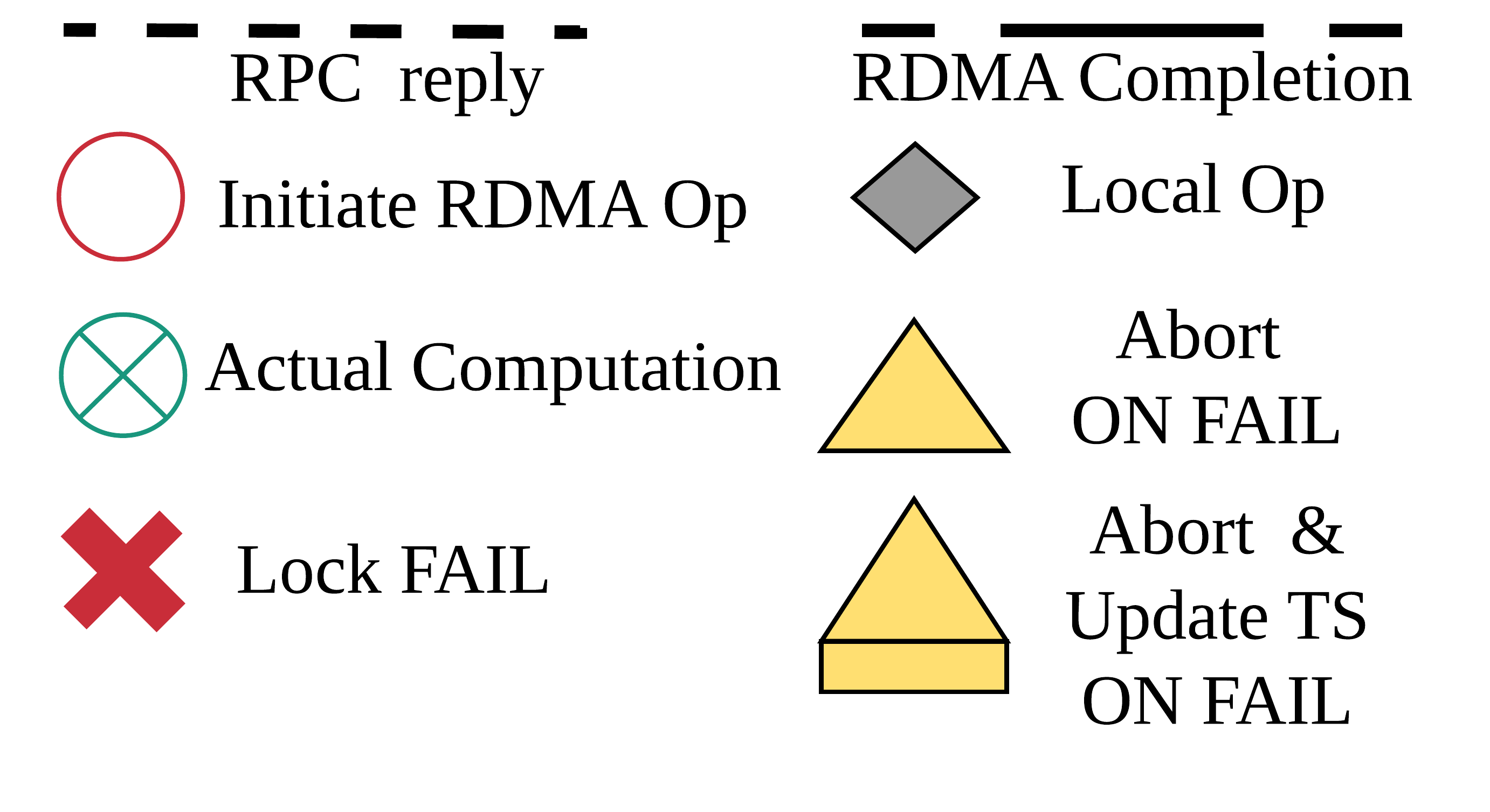}
    \vspace{-0.4cm}
    \caption{Legend used for this section 
    %\red{PUT IN A TABLE}
    }
     \vspace{-0.4cm}
    \label{fig:legend}
% \end{wrapfigure}
\end{figure}

Next, we describe the implementation of each protocol in \projectname
%the step-by-step implementations of \projectname's concurrency control protocols 
except for \occ, which
is implemented based on DrTM+H~\cite{wei2018deconstructing}. 
We choose to base our \occ implementation on DrTM+H because it outperforms other RDMA-based \occ implementations by \cite{dragojevic2014farm} and \cite{kalia2016fasst}. 
Figure \ref{fig:legend} shows the legend for protocol operations in this section. 
% \red{==> PLEASE CHANGE TO A TABLE, SEE HOW TO PLACE THE LEGENDS IN THE TABLE..
% } 

%used throughout this section. We leverage the logging functionality directly from DrTM+H~\cite{wei2018deconstructing}.
%to support two types of logging and 2PC across \projectname: two-sided RPC and one-sided \texttt{WRITE}. 
%To highlight protocol-specific operations, we omit common logging and 2PC in the figure for each protocol. 
%For other operations, We try our best to give detailed descriptions so that the paper can serve as a reference for others to implement the protocols in RDMA. No prior work has shown the comprehensive designs, and we will also open-source our framework if the paper is accepted. 

\subsection{NOWAIT}
% \vspace{-2mm}

% \setlength{\intextsep}{2pt}%
% \setlength{\columnsep}{8pt}%
% \begin{wrapfigure}[13]{r}{0.7\linewidth}

\begin{figure}[htp]
    \centering
    \vspace{-0.2cm}
    \includegraphics[width=0.8\linewidth]{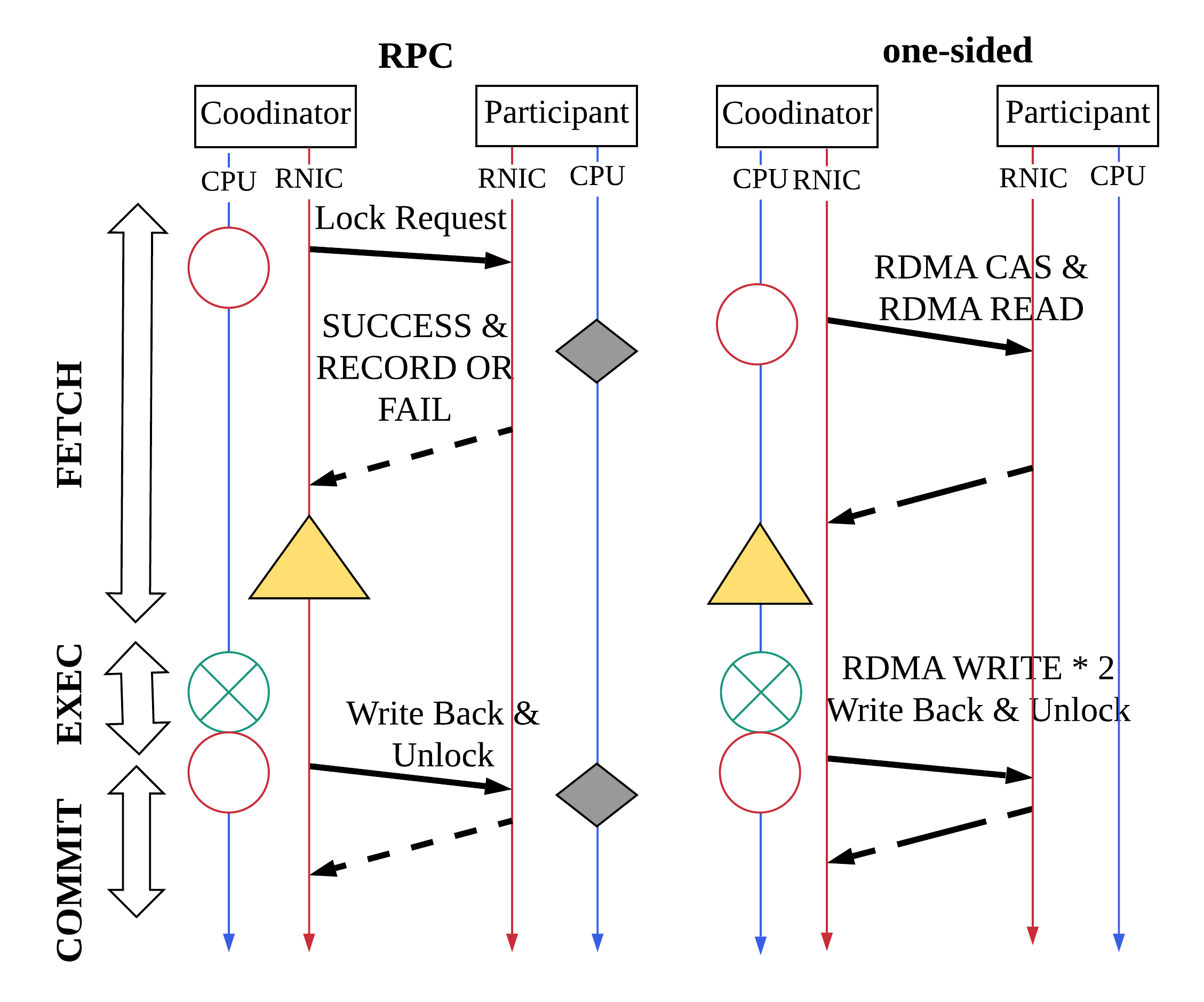}
    \vspace{-4mm}
    \caption{\nowait Implementations}
     \vspace{2mm}
    \label{fig:nowait}
% \end{wrapfigure}
\end{figure}
 
\nowait~\cite{Bernstein:1981:CCD:356842.356846} is a basic concurrency control algorithm 
based on 2PL that prevents deadlocks. 
%is the most basic way of deadlock prevention mechanism among 2-phase locking concurrent control algorithms. 
%It was proposed by ~\cite{Bernstein:1981:CCD:356842.356846} in the 1980s. 
A transaction in \nowait tries to lock all the records accessed; if it fails to lock a record that is already locked by another transaction, 
the transaction aborts immediately and releases all locks that have been successfully acquired. Figure~\ref{fig:nowait} shows the operations of \nowait for both RPC and one-sided implementations.

%In \nowait, lock and commit are two 
%operations relevant to RDMA implementation. 
%to prevent deadlock. 
%Figure~\ref{fig:nowait} shows our RPC and one-sided implementations of \nowait. 

With RPC, a coordinator locks records by sending RPC locking request
to the corresponding participant, 
the RPC handler locks the record using a local CAS.
If the CAS fails, a failure message is sent 
back to the coordinator which will
release all read and write
locks by posting RPC release requests before aborting the transaction.
Otherwise, the participant's handler has already locked the tuple locally, and it returns a success message with the record in response.
On transaction commit, with all locks acquired,
a write-back request associated with
the updated records is sent to each participants,
where an RPC handler 
performs write-back of the record and releases the lock.

\begin{comment}
{\bf RPC}
%In the RPC implementation,
%is straight forward. As shown in Figure~\ref{fig:nowait}, 
In the data fetch stage, a coordinator locks records by sending RPC locking request
to the corresponding participant (\step \textbf{a1}).
%will trigger the send of an RPC request for locking
%to the corresponding participant (\step \textbf{a1}).
%sending locking a RPC request to each corresponding participant (\step \textbf{a1}). 
An RPC handler is called to lock the record using local CAS (\step \textbf{a2}). If the CAS fails, a failure message is sent back, and the coordinator will release all read and write
locks by posting RPC release requests before aborting the transaction (\step \textbf{a3}).
%\red{WHAT OPERATION DO WE USE FOR 
%RELEASING LOCK? ONE-SIDED?}. 
Otherwise, the participant's handler has already found and locked the record locally, and it then returns a success message with the actual record as an RPC reply (\step \textbf{a2}). Upon all records are collected, the transaction runs locally at the coordinator side.
At the commit stage, 
a write-back request with the updated record 
is sent back to each participant (\step \textbf{a4}), where an RPC handler 
performs write-back of the record and releases the lock (\step \textbf{a5}).
\end{comment}

With one-sided primitive, 
we use the doorbell batching mechanism as
an efficient way to issue multiple outstanding requests from the sender. 
With this optimization, only one yield is needed after the last request is posted, thus reducing latency and context switching overhead.
On locking, the coordinator needs to perform two 
operations---RDMA \texttt{CAS} and \texttt{READ}---to lock and read the 
remote record. 
Logically, they should be performed one
after another, but in fact, the coordinator
can issue \texttt{READ} immediately
after \texttt{CAS} to 
overlap the communication. 
It is because the two will be performed 
in the issue order remotely, and if the lock 
acquire fails, the coordinator can 
simply ignore the returned data of \texttt{READ}.
Note that the read offsets are collected and cached by the coordinator 
before transaction execution starts and 
do not incur much overhead.
With high contention, the optimization
tends to add wasted network traffic.  
However, for network-intensive applications with low contention, i.e., SmallBank, the throughput increases by 25.1\% while average latency decreases by 22.7\%.
Similarly, two RDMA \texttt{WRITE}s are posted to update and unlock the record at the 
commit stage.
Only the second RDMA write is signaled to avoid sending multiple MMIOs and wasting PCIe bandwidth. 
Different from lock \& read, 
the doorbell batched update \& unlock is always beneficial.
\subsection{WAITDIE}
% \vspace{-2mm}

Different from \nowait, which 
unconditionally aborts any transaction accessing conflicting records, \waitdie
resolves conflicts with a global consensus priority.
On start, each transaction obtains a globally unique timestamp, which can be stored in 
the lock records it accessed. 
Upon detecting a conflict, the timestamp logged
in the lock is compared with the current 
transaction's timestamp to determine whether 
to immediately abort 
the transaction or let it wait. 
In \projectname, we construct the 
unique timestamp of a transaction by appending
the machine ID, thread ID, and coroutine ID
to the low-order bits of the local clock time~\cite{bernstein1983multiversion}.
This avoids the high overhead of global clock synchronization 
such as such as NTP~\cite{10.1145/956981.956983}
and PTP~\cite{lee2005ieee}.
The timestamp can be stored in the 64-bit lock record. 

\begin{figure}[htp]
    \centering
    \vspace{-0.2cm}
    \includegraphics[width=0.8\columnwidth]{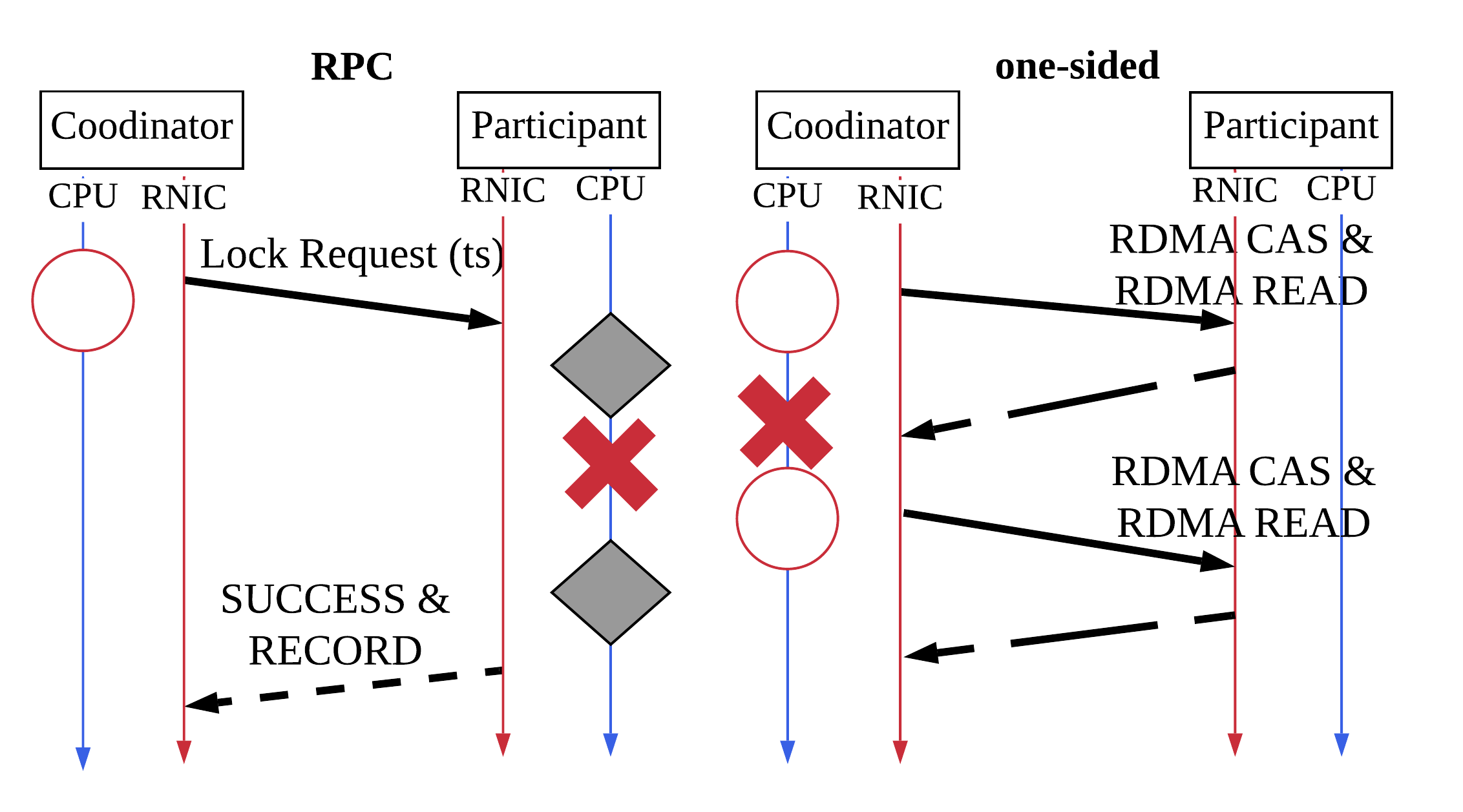}
    \vspace{-0.6cm}
    \caption{\waitdie: the \textbf{FETCH} stage
    %: the locking phase when the requesting transaction needs to WAIT 
    %instead of DIE.
    }
    \vspace{0cm}
    \label{fig:waitdie}
% \end{wrapfigure}
\end{figure}

Compared to \nowait,
the new operation in \waitdie is transaction wait.
Figure~\ref{fig:waitdie} shows the \waitdie operations in the fetch stage. With RPC, it can be implemented easily:
when an accessed record is locked, the lock request
handler can decide
based on the request transaction's timestamp 
whether to let it wait until it is unlocked, 
or send back a failure reply immediately.
Note that the handler does not 
busy wait for the lock 
on behalf of the transaction, and block other incoming requests. 
Instead, the transaction is added to the lock's waiting list, which is checked in the event loop periodically by the handler thread. On lock release, the handler thread removes the transaction from the waiting list and replies to the coordinator with a success message and the locked record.

% \setlength{\intextsep}{2pt}%
% \setlength{\columnsep}{8pt}%
% \begin{wrapfigure}[11]{r}{.51\linewidth}

%{\bf RPC} When an accessed record is locked, the lock request
%handler decides
%based on the request's timestamp 
%whether to wait on the record until it is unlocked 
%or send back a failure reply immediately (\step \textbf{a2}).
%encountering a locked record, 
%based on the request's timestamp, the lock request handler will decide whether to wait on the record until the record is lockable or send back a failure reply immediately (\step \textbf{a2}).
%While waiting for a locked record to be released,
%To wait for the record to unlock, 

%This design ensures the balance between throughput and latency. 

With one-sided primitive, the implementation 
is less straightforward.
The key difference is that the current transaction 
needs to obtain the record's timestamp---even if it is locked---and decides to abort or wait by itself. 
Similar to \nowait, we use an RDMA \texttt{CAS} 
followed by an RDMA \texttt{READ} to 
retrieve the remote lock together with
its timestamp and record, as seen in Figure~\ref{fig:waitdie}.
If the record is not locked, the CAS succeeds and atomically writes the transaction's timestamp on the remote lock, and returns 0.
If the CAS fails,i.e.,the record is locked, 
rather than abort immediately, 
the current transaction compares its timestamp with 
the returned timestamp, which indicates the
lock-holding transaction,
to determine whether to abort itself or wait.
If the decision is to wait, the co-routine keeps posting RDMA \texttt{CAS} with \texttt{READ} requests and yields after every unsuccessful trial until it succeeds.

{\bf Limitation} Current one-sided implementation of \waitdie is not starvation-free for old transactions: when the oldest transaction fails to lock, the lock may be released and reacquired by another younger transaction, making the oldest transaction starve. One potential solution may be that a counter is put along with the timestamp and initialized to be 0. When an old transaction detects failure after the first \texttt{CAS} \& \texttt{Read}, it increments the counter once by issuing an RDMA \texttt{
FETCH\_AND\_ADD} operation, all future younger transactions accessing the record will then abort until the counter is reset to 0. Another 
\texttt{FETCH\_AND\_ADD} is needed to decrement the counter when the old transaction successfully grabs the lock.
\subsection{MVCC}
% \vspace{-2mm}

\begin{comment}
\begin{figure}[t]
    \begin{minipage}{.49\linewidth}
    \centering
    \includegraphics[width=\linewidth]{images/MVCC-write.pdf}
    \vspace{-10mm}
    \caption{\mvcc: Write}
    \label{fig:mvcc-write}
    \end{minipage}
    %\quad
    \begin{minipage}{.49\linewidth}
    \centering
    \includegraphics[width=\linewidth]{RCC/images/MVCC-read_no_number.pdf}
    \vspace{-10mm}
    \caption{\mvcc: Read}
    \label{fig:mvcc-read}
    \end{minipage}
    \vspace{-8mm}
\end{figure}
\end{comment}

\begin{figure}[htp]
    \centering
    \vspace{-0.3cm}
    \includegraphics[width=0.8\columnwidth]{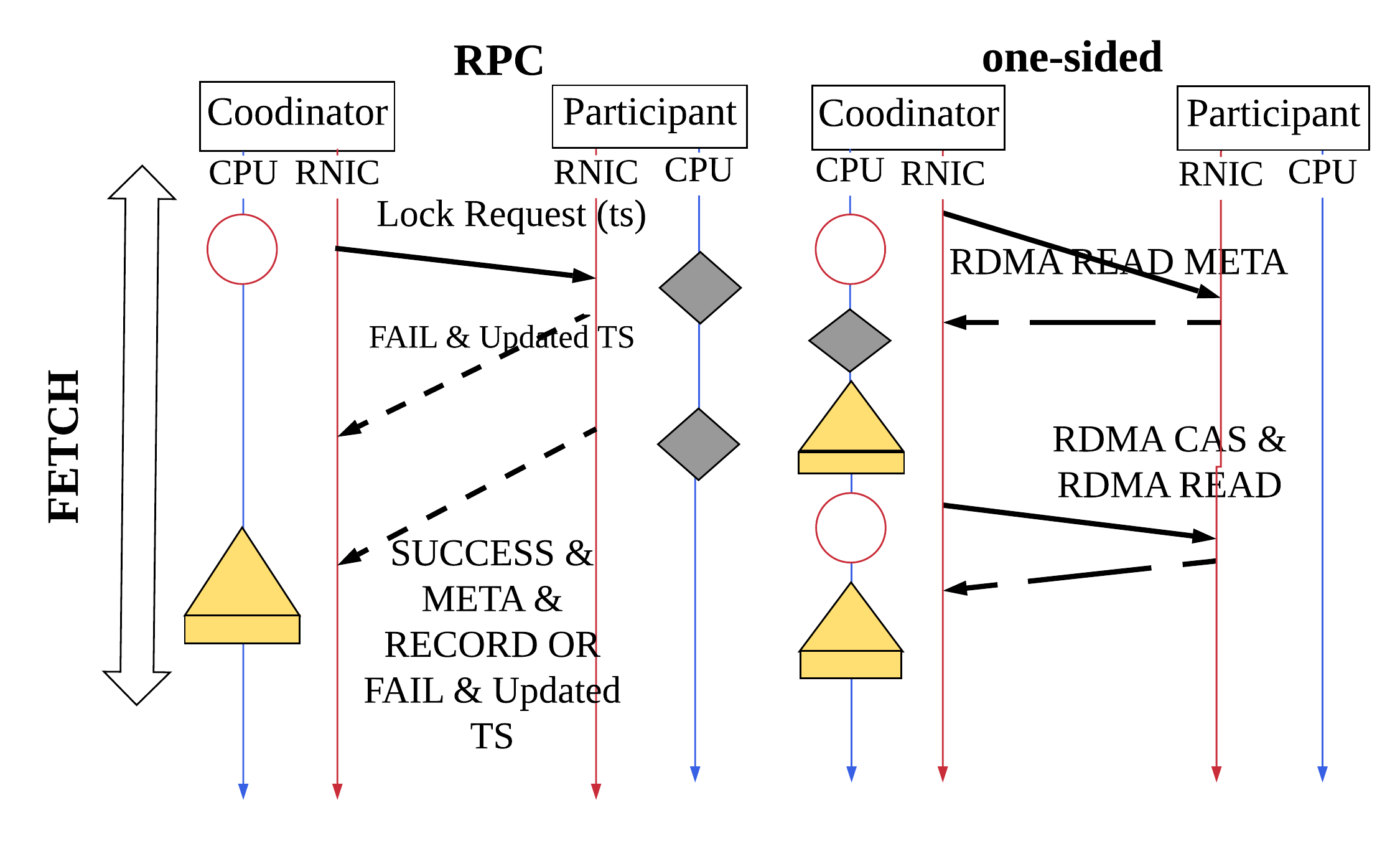}
    \vspace{-0.6cm}
    \caption{\mvcc: Write}
    % \vspace{-0.4cm}
    \label{fig:mvcc-write}
\end{figure}

\begin{figure}[htp]
    \centering
    \vspace{-0.2cm}
    \includegraphics[width=0.8\columnwidth]{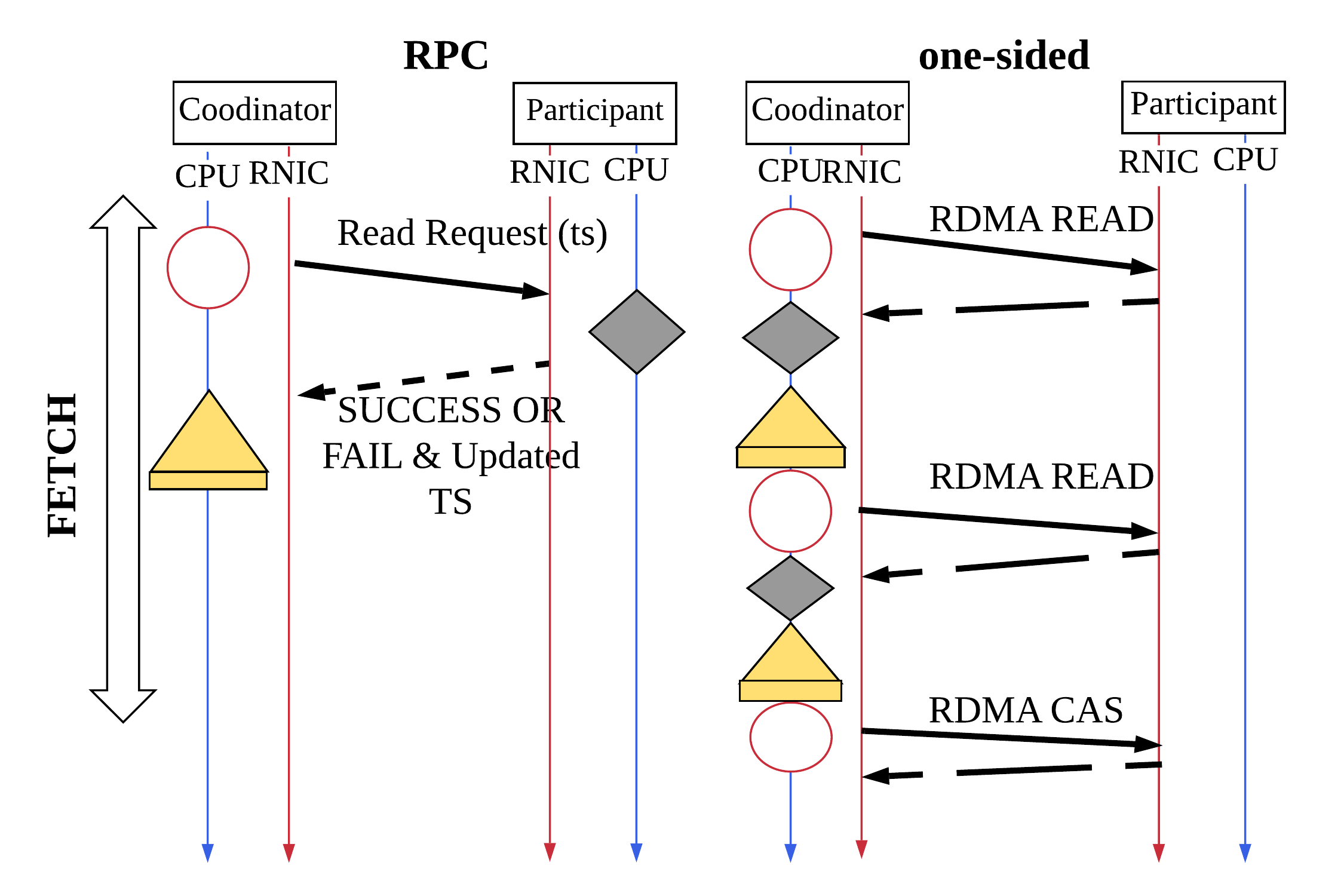}
    \vspace{-0.4cm}
    \caption{\mvcc: Read}
    \vspace{-0.4cm}
    \label{fig:mvcc-read}
\end{figure}

\mvcc (Multi-Version Concurrent Control)~\cite{bernstein1983multiversion} 
reduces read-write conflicts 
by keeping multiple versions of the record 
and providing a recently committed version when possible.
Shown in Figure~\ref{fig:metadata},
the metadata of each tuple in 
\mvcc consists of three parts:
1) write lock, which contains the timestamp of the current transaction holding the lock
that has not committed yet (\texttt{tuple.tts});
2) read timestamp (\texttt{tuple.rts}), which is the latest (largest) transaction timestamp that has successfully read the record; and
3) write timestamps (\texttt{tuple.wts}), which are  
the timestamps of recently {\em committed} transactions that have performed writes on the record. These versions are kept in the participants.
We also denote the timestamp of the current transaction trying to fetch records as \texttt{ctts}.

To access a record in RS, 
we check 
{\em Cond R1:} 
there is a proper record version
based on the \texttt{tuple.wts} of recently committed transactions---it should
choose the largest \texttt{tuple.wts} smaller than \texttt{ctts}; and
{\em Cond R2:} \texttt{tuple.tts} is 0 or larger than \texttt{ctts}.
Cond R2 means there is no un-committed transaction writing
the record, or the write happens {\em after} the read, in which the read transaction 
can still correctly gets one of the committed versions of the record.
If both Cond R1 and R2 are satisfied, the version from Cond R2 can be returned.

To access a record in WS, we check
%A write can be only performed if 
{\em Cond W1:} transaction's timestamp is larger than 
the maximum \texttt{tuple.wts} and the current \texttt{tuple.rts}; and 
{\em Cond W2:} the record is not locked.
If either is failed, the transaction is aborted; otherwise,
the record is locked with \texttt{tuple.tts} updated to \texttt{ctts},
a new record is created and
sent back to the transaction. 
%Note that \texttt{tts} is not the same as any \texttt{wts}, which
%indicates the timestamp of a committed transaction.

Conceptually, \mvcc maintains the following
properties.
A write of transaction \texttt{ctts} 
cannot be ``inserted'' among the committed transactions indicated by \texttt{tuple.wts}; 
and the write should be ordered after any performed read.
A read should always return the most recent committed version of a record. 
The key requirement for correctness is that 
the condition check for RS and WS record
should be {\em atomic}. 

The original \mvcc requires using a linked list to 
maintain a set of record versions. 
%\red{With enough versions, a read can always find the 
%suitable version to return.}
However, the nature of one-sided primitive makes it 
costly to traverse a remote linked list---in the worst case,
the number of one-sided operations for a single remote
read is proportional to the number of versions
in the list.
Thus, we use a static number of memory slots allocated for each record to store the stale versions.
A transaction will simply abort when it cannot find a suitable version among the slots available for a read operation.
The number of slots determines the trade-off between 
the extra read aborts and 
reduced memory/traversal overhead.
We choose four slots because 
our preliminary experiments show that
at most 4.2\% of read aborts are 
due to slot overflow.

In \mvcc, we use the same timestamp organization as \waitdie.
The local clock reduces bandwidth overhead of a global clock but may introduce significant bias.
{While not affecting correctness,
the large time gap between different machines may lead to a long waiting time.
To mitigate the issue, each transaction co-routine maintains a local time and {\em adjusts} the 
local time whenever it finds a larger \texttt{tuple.wts} or \texttt{tuple.rts}
in any tuple received.
The encapsulated remote time on the \texttt{tuple.wts} or \texttt{tuple.rts} is extracted and local time is adjusted accordingly if the extracted remote time is larger.}
This mechanism limits the gap of local timer between machines, and 
reduces the chance of abort due to the lack of suitable version among the fixed version slots. 

While it is not hard to conceptually understand \mvcc,
the implementation with RDMA needs to 
ensure atomicity. 
Let us first consider accessing records in WS. 
One way is to first check Cond W2 and lock the record, 
at this point, the metadata cannot be accessed by other writes,
we can reliably check Cond W1. If it is not satisfied, 
the lock is released and the write transaction aborts.
However, in this way we need to perform a lock for every write, even if the 
write transaction cannot be properly serialized.
It is particularly a problem for one-sided primitives, 
because the lock is implemented with an RDMA \texttt{ATOMIC CAS}.
The better approach is to first check Cond W2 and then acquire the lock. 
However, a subtle issue raises because
Cond W1 and Cond W2 are not done atomically. 
Between the point that Cond W1 is satisfied and 
the point the lock is acquired,
another transaction that writes the record 
can lock the record and commit (unlock).
According to the protocol property, the current transaction should 
be aborted, but it will find both Cond W1 and Cond W2 satisfied. 
To ensure atomicity while avoiding 
the overhead of locking. We propose the 
{\em double-read} mechanism.
After the lock is acquired, 
Cond W1 should be checked {\em again}, if it is still satisfied,
the write can proceed, otherwise, it is aborted.

As in Figure~\ref{fig:mvcc-write}, with RPC, the write protocol can be implemented by the handler on the participant. 
With one-sided primitive, 
the coordinator posts an RDMA \texttt{READ} to read the 
metadata of the record---\texttt{tuple.rts} and \texttt{tuple.wts}---on the participant, then
checks Cond W1 locally.
If it is satisfied, the coordinator posts an RDMA \texttt{ATOMIC CAS}
to lock the record, and a second RDMA \texttt{READ} to fetch the tuple.
Cond W1 can be checked again based on the 
just returned \texttt{tuple.rts} and \texttt{tuple.wts},
if it still holds, the returned record \texttt{tuple.record} is kept locally in 
the coordinator. Otherwise, the transaction aborts
and the lock on the record is released.
%Remember that the write operation handles records
%in both read and write set, so the program logic that 
%triggers the write procedure can be actually a read. 
%This is why the record should 
%be returned by the second RDMA \texttt{READ}.

\begin{comment}

%Then, in \step \textbf{b2}, the received data are checked locally.
A write operation can only succeed if the transaction's \texttt{tts}
is larger than the received \texttt{rts} and all \texttt{wts} in 
the tuple (same as in RPC).
%in the transaction can only be valid if the tts is larger than the rts and any of wts on the tuple. 
If the check fails, the coordinator updates its \texttt{tts} to the largest timestamp and abort. 
Otherwise, the coordinator posts RDMA \texttt{ATOMIC CAS} and RDMA \texttt{READ} to lock the remote record and fetches the metadata and record in \textbf{b4}, similar to \nowait and \waitdie. 
If \texttt{ATOMIC CAS} fails, the transaction aborts and updates its \texttt{tts} if the timestamp on the lock is larger. 

\end{comment}

When accessing records in RS, 
the tuples need to be fetched atomically.
%The read is also complicated by the non-atomic
%read of metadata and record. 
%For a read in current transaction with \texttt{ctts}, 
%it can generate a version by checking Cond 1R and 2R. 
%However, between the access of \texttt{wts} and the record,
%another transaction can be committed, changing \texttt{wts}.
%It can affect correctness because the read may need to return
%the version of the just committed transaction, but it was not in
%the originally read \texttt{wts}.
The separate double-read mechanism
discussed before can be generalized to 
{\em two consecutive reads of the same tuple}. 
If the contents of each returned data are the same, 
then we are sure that atomicity is not violated.
%fetching the metadata again, and check whether \texttt{wts} is 
%the same as before.
Based on the atomically read
tuple, Cone R1 and Cond R2 can be checked 
to generate the appropriate committed
version for the record in RS.
If the second tuple returned is different
from the first, then the transaction is simply 
aborted. 
We apply a small optimization to reduce unnecessary
abort: among the two versions of metadata,
we only need to ensure the match of
\texttt{tuple.wts}.
The \texttt{tuple.tts} can be different since
a transaction corresponds to the first \texttt{tuple.tts} can be aborted between the 
two reads. But as long as Cond R2 is satisfied, 
the read can still get a version among 
\texttt{tuple.wts}.
%If it is true, no new transaction has committed and we can 
%correctly return the record from the first read.
%Otherwise, the read transaction is aborted.
%Note that 
%for the second time, 
%we do not need to check Cond 2R:
%if the original transaction indicated by \texttt{tts} of the record
%already committed, the new one (if the second \texttt{tts} is non-zero)
%must have a timestamp even larger, so the read record can also be returned.

As in Figure~\ref{fig:mvcc-read}, with RPC, the read procedure can be implemented
in a straightforward manner with the handler on participant.
With one-sided primitives, the two reads are implemented by 
two doorbell batched RDMA \texttt{READ}s.
The only additional operation is
to use an RDMA \texttt{ATOMIC CAS} to update \texttt{rts} of the 
record in the participant. If it fails, we can simply retry until 
succeed. Note that it does not imply conflict, but just multiple 
concurrent reads.

On commit, with one-sided primitive, 
the coordinator locally overwrites the oldest \texttt{wts}
with its own \texttt{ctts}, and updates 
the corresponding record to the locally
created one for write. 
Then it posts two RDMA \texttt{WRITE}s.
The first write puts the locally prepared new record+metadata
to the participant; the second write releases the lock.
With RPC, the procedure can be implemented similarly. 

\textbf{Garbage collection \& memory management} Since our \mvcc uses a static number of slots instead of employing a linked list, all slots are pre-allocated both for the use of two-sided RPC function calls and for one-sided RDMA access. Therefore, it is unnecessary to garbage collect stale versions when they are out of visibility of any read/write.

\textbf{Clock synchronization} \mvcc uses local clock plus adjustment instead of global clock synchronization to avoid wasting network bandwidth. Global synchronization protocols like NTP~\cite{10.1145/956981.956983} are typically used to keep machines synchronized with the Internet within milliseconds skew. PTP~\cite{lee2005ieee} can synchronize network computers within sub-milliseconds skew by employing a 
Best Master Clock (BMC) algorithm. Our synchronization technique integrates the adjustment within the \mvcc protocol, making the adjustment on demand.

\subsection{SUNDIAL}
% \vspace{-2mm}

\begin{comment}
\begin{figure}[t]
    \begin{minipage}{.49\linewidth}
    \centering
    \includegraphics[width=\linewidth]{images/Sundial-write.pdf}
    \vspace{-12mm}
    \caption{\sundial: Write}
    \label{fig:sundial-write}
    \end{minipage}
    %\quad
    \begin{minipage}{.49\linewidth}
    \centering
    \includegraphics[width=\linewidth]{images/Sundial-read.pdf}
    \vspace{-12mm}
    \caption{\sundial: Read}
    \label{fig:sundial-read}
    \end{minipage}
    \vspace{-4mm}
\end{figure}
\end{comment}

\begin{figure}[htp]
    \centering
    \vspace{-0.3cm}
    \includegraphics[width=0.8\columnwidth]{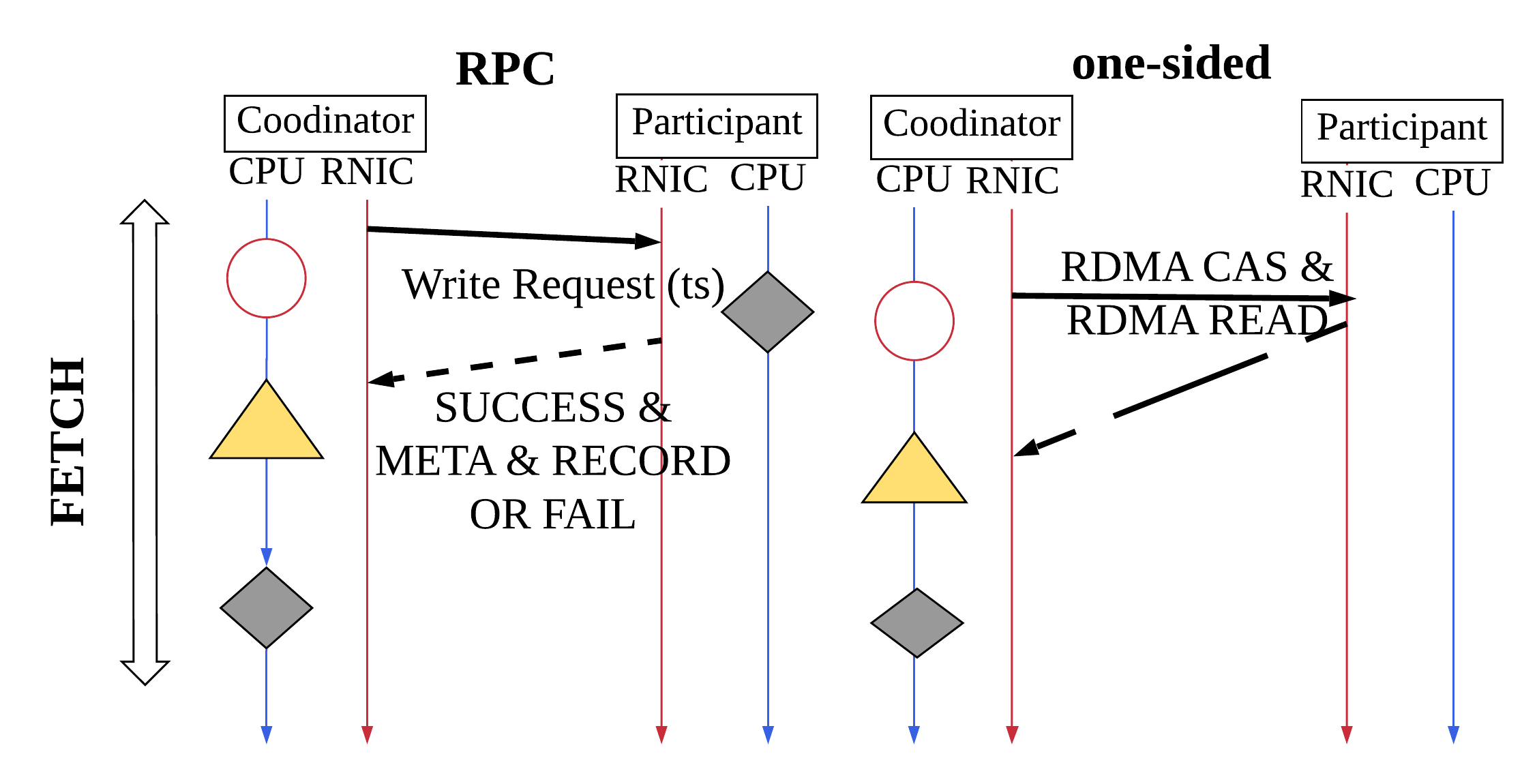}
    \vspace{-0.4cm}
    \caption{\sundial: Write}
    % \vspace{-0.4cm}
    \label{fig:sundial-write}
\end{figure}

\begin{figure}[htp]
    \centering
    \vspace{-0.2cm}
    \includegraphics[width=0.8\columnwidth]{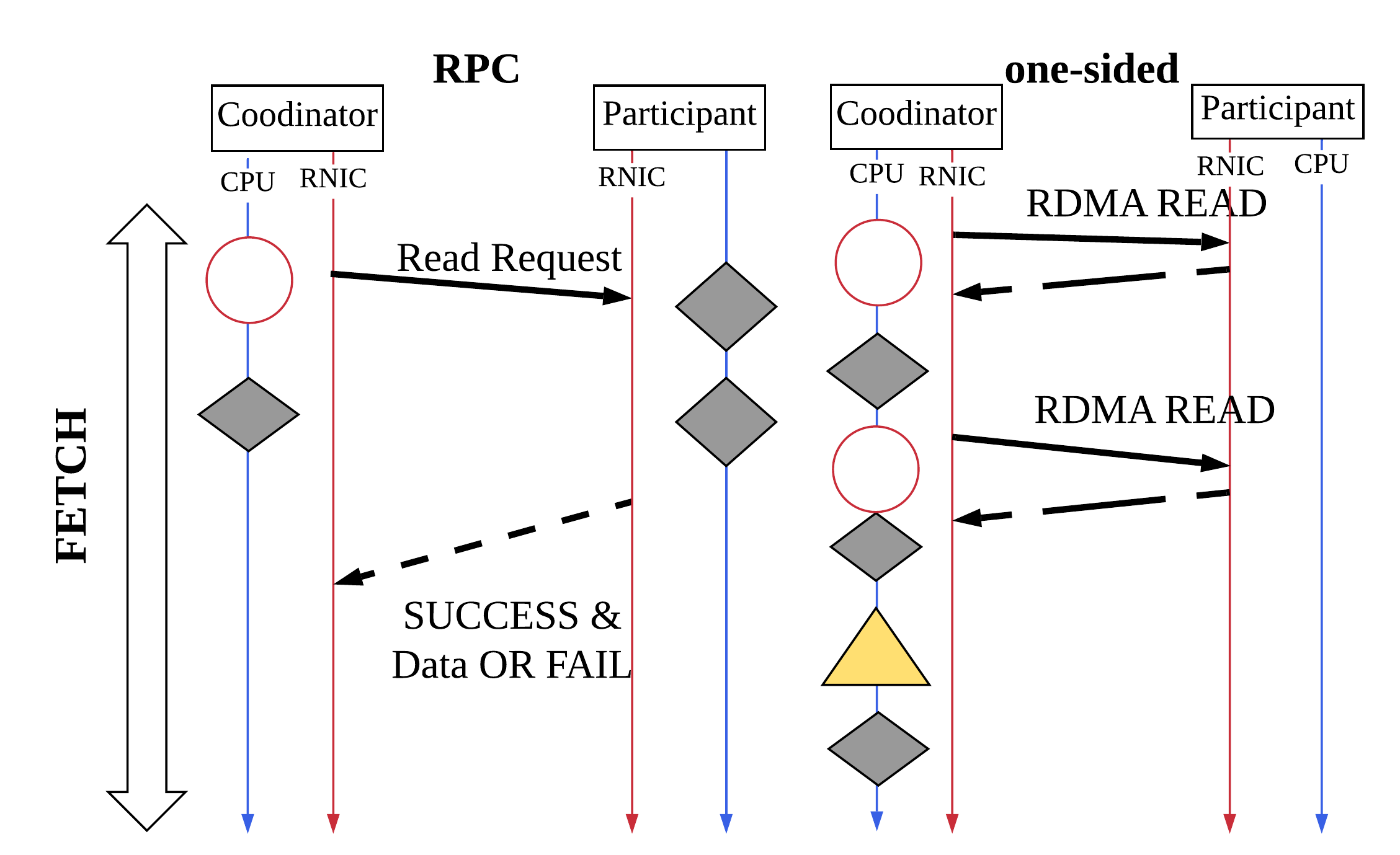}
    \vspace{-0.4cm}
    \caption{\sundial: Read}
    % \vspace{-0.2cm}
    \label{fig:sundial-read}
\end{figure}

\sundial~\cite{yu2018sundial} 
is an elegant protocol based on logical leases to avoid unnecessary
aborts while still maintaining serialization by 
dynamically adjusting the timestamp of transactions or commit order. 
Based on the tuple format in Figure~\ref{fig:metadata}, 
the lease of a tuple is specified by \texttt{tuple.[wts,rts]}.
Each transaction has a \texttt{commit\_tts}, which indicates the 
{\em required} timestamp of the transaction to satisfy the current 
lease of accessed records.
When accessing a record in RS, the transaction {\em atomically}
reads the tuple and update \texttt{commit\_tts} to 
\texttt{Max(commit\_tts,tuple.wts}).
It is because to correctly read the record, the transaction has
to be logically ordered after the most recent writer transaction. 
When accessing a record in WS, the transaction 
tries to lock the tuple, and if it is also in RS, checks
whether \texttt{tuple.wts} is the same as the \texttt{RS[key].wts}.
The second condition ensures that there is no transaction 
writing the record committed since the read. 
If both conditions pass, the transaction's \texttt{commit\_tts} 
is updated to \texttt{Max(commit\_tts,tuple.rts+1)}.
It ensures that the transaction is logically ordered after the 
current lease of the record. Since other transactions may have read
the record during the lease, without such update, the transaction
would have to be aborted. 

Although the update of \texttt{commit\_tts} during execution 
will try to satisfy the {\em current} lease based on {\em individual}
record, at the commit time, the transaction needs to be validated to ensure
its current \texttt{commit\_tts} falls into {\em all} leases
of records in RS.
If it is not satisfied, \sundial allows the transaction 
attempt to {\em renew} the lease by adjusting the \texttt{tuple.rts}
in the data store at participant~\footnote{The condition 
\texttt{commit\_tts} must be greater than \texttt{wts} of the 
record in RS based on how it is updated}.
The renew is failed if
(1) the current \texttt{wts} is not the same as current \texttt{tuple.wts}, meaning that there is a later committed transaction writing the record, 
which invalidate the previous read record; or
(2) the record is locked, meaning that there is a transaction 
trying to write the record, which prevents the lease extension.
Otherwise, the transaction can adjust the 
lease by updating \texttt{tuple.rts} to \texttt{commit\_tts}.
The key requirement is that the lease renewal operation should be
performed {\em atomically}.
If all RS records are validated, and all necessary
lease renewals are successful, the transaction
is committed, which updates \texttt{tuple.wts}
and \texttt{tuple.rts} of all records in WS
to be \texttt{commit\_tts}.

For records in WS with one-sided primitives, the tuples can be easily checked after a doorbell batched \texttt{CAS} and \texttt{READ} to lock and retrieve the tuple, as in Figure~\ref{fig:sundial-write}. Yet to implement \sundial in \projectname, we need to 
solve two problems. 
First, for records in RS, the tuple needs to be accessed atomically.
This can be done using the double doorbell batched reads with 
one-sided primitives or simply double read with RPC
introduced in \mvcc, as shown in Figure~\ref{fig:sundial-read}.
Second, we need to ensure the atomic lease renewal, which is more
challenging than atomic read. 
To implement this, we first atomically read the tuple from participant, 
then use an atomic operation to update \texttt{tuple.rts}.
With these two ideas, we can implement 
RPC and one-sided version of \sundial.

In RPC version, the atomic tuple read and lease renewal are 
all performed by the handler in the participant. 
The coordinator just poses the read and renewal requests and
processes the responses according to the protocol.
In one-sided version, the fetch of tuples in RS and WS
is similar to \mvcc with double doorbell batched reads.
Based on the fetched tuples, the coordinator locally performs
the \sundial protocol operations. 
For lease renewal, the coordinator first atomically reads the tuple,
then checks the lease extension condition, if it is allowed, 
it poses an RDMA \texttt{ATOMIC CAS} with the previous \texttt{tuple.rts}
is the old value and its \texttt{commit\_tts} as the new value.
In this way, the lease renewal is performed atomically.
It is worth noting that we can implement in this manner
because the \sundial protocol only requires updating one variable
\texttt{tuple.rts} to renew the lease. If multiple variables
need to be updated, then more sophisticated mechanisms 
are needed and it is beyond the scope of the paper.

\subsection{CALVIN}
% \vspace{-2mm}

\begin{comment}
\setlength{\intextsep}{2pt}%
\setlength{\columnsep}{8pt}%
% \begin{wrapfigure}[9]{r}{.8\linewidth}
\begin{figure}[htp]
    \centering
    \vspace{-0.4cm}
    \includegraphics[width=0.9\linewidth]{images/CALVIN-overview-2.pdf}
    \vspace{-5mm}
    \caption{\calvin threading model}
    % \vspace{-0.6cm}
    \label{fig:calvin-overview}
% \end{wrapfigure}
\end{figure}
\end{comment}

Different than all other protocols, 
\calvin~\cite{Thomson:2012:CFD:2213836.2213838} enforces a deterministic 
order among transaction in an epoch, e.g., all transactions received by the 
system during a certain time period.
The readers can reference the original paper 
for the complete motivation and advantages of this approach, we are interested
in how the communication happen and can be implemented in RDMA for such a protocol.

\begin{figure}[htp]
    \centering
    % \vspace{-0.2cm}
    \includegraphics[width=0.9\columnwidth]{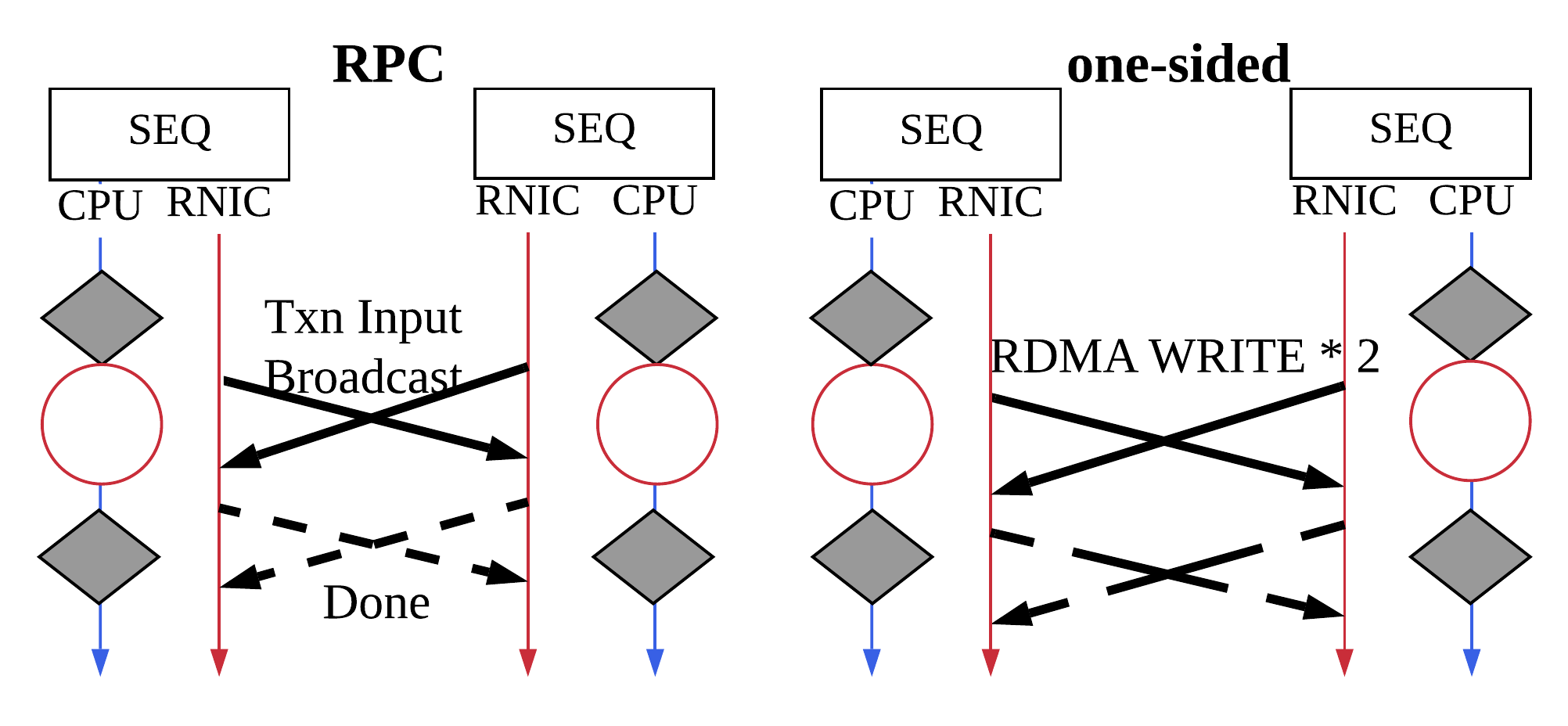}
    \vspace{-0.4cm}
    \caption{\calvin: Txn Input Broadcasting}
    %  \vspace{-0.6cm}
    \label{fig:calvin-det-framework}
\end{figure}

In \projectname, \calvin works as follows. 
For each epoch, each machine node receives a set of transactions. 
The sequencing layer in each machine determines the order of the locally
received transactions and broadcasts them to all other machines. 
After the transaction dispatch, each machine has the whole set
of transactions in the epoch with a consensus and deterministic order.  
%Also, based on certain static rules, the order of all transactions 
%is determined and then each machine should execute them according this order.
The transaction dispatch incurs \calvin's first source of communication:
the transaction inputs, its RS and WS, will be delivered to all 
other machines. 
With RPC, such information can be sent in batch and the receiver nodes
will store the data locally. 
With one-sided primitives, the implementation is more challenging, 
since the sender node needs to be aware of the location to write to 
remote nodes. We design a specific buffer structure,
in each node that is known among all machines, so that the sender
can directly use doorbell batched RDMA \texttt{WRITE}s to deliver the transaction information to all other nodes and update metadata, as in Figure~\ref{fig:calvin-det-framework}.

\begin{figure}[ht]
    \centering
    % \vspace{-0.2cm}
    \includegraphics[width=0.8\linewidth]{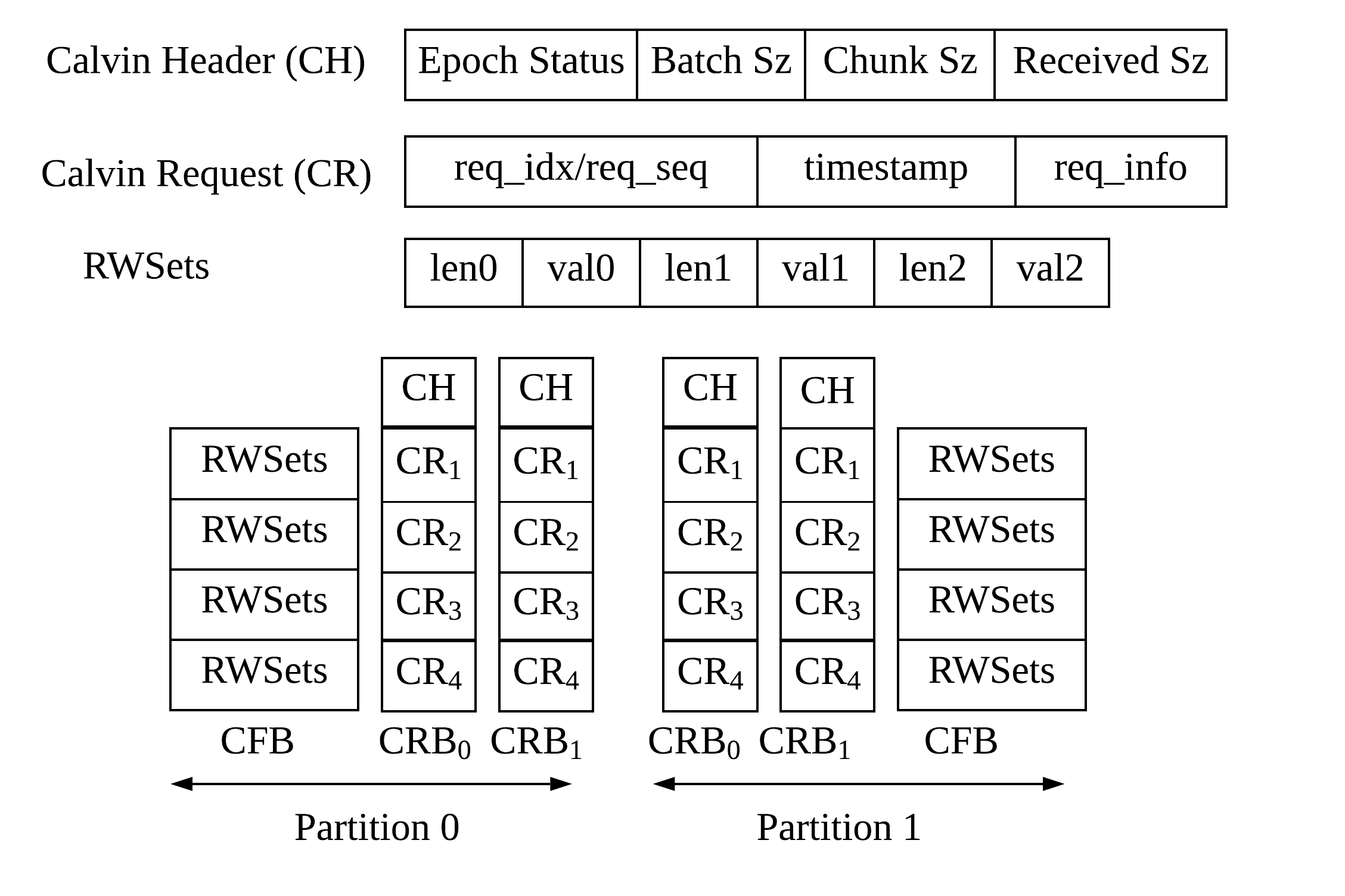}
    \vspace{-0.4cm}
    \caption{RDMA-enabled buffer organization with batch size per epoch = 4 and the maximum number of read/write sets supported per transaction = 3.}
    \vspace{0cm}
    \label{fig:calvin-buffers}
% \end{wrapfigure}
\end{figure}

Figure~\ref{fig:calvin-buffers} exemplifies the buffer organization design for \calvin. \projectname \calvin uses two memory buffers that enable RDMA remote access. %Figure~\ref{fig:calvin-buffers} shows an example buffer organization. 
1. \calvin Request Buffer (CRB). Each CRB contains one \calvin Header (CH) and a list of \calvin Requests (CR). Each CH has control information for \calvin's scheduler to decide whether it has collected all transaction inputs in one epoch and whether all transactions in a batch have finished execution and all threads should move on to the next epoch. 2. \calvin Forward Buffer (CFB). Each execution co-routine uses one CFB to receive forwarded values from other machines. We will discuss \calvin's value forwarding in later paragraphs. Besides these two buffers above, to support asynchronous replication, each backup machine has a list of CRBs for receiving asynchronous backup requests for each epoch.

\setlength{\intextsep}{2pt}%
\setlength{\columnsep}{8pt}%
% \begin{wrapfigure}[11]{r}{.48\linewidth}

\calvin has the unique execution model that each transaction is 
executed by multiple machines. 
Specifically, the machines that have records in WS according to
data partition will 
execute the operations of the transaction that will write these records.
These machines are called active participants. 
The machines that have records in RS are called passive 
participants, since they do not contain records in WS, they do not 
execute the transaction but only provide data to active participants. 
To start the execution in active participants, 
they need to get the complete set of records in RS and WS. 
This leads to the second source of communication in \calvin. 

First, the passive participants need to send the local records in RS to all active participants. 
Second, the active participants need to send the local records in WS
to the other active participants. Actively participants will wait and collect all the needed records forwarded from other machines. Two-sided implementations is easier since we can simply use a data structure for holding the mappings from tuple key to their values in the epoch. The one-sided version needs two doorbell-batched RDMA \texttt{WRITE}s to forward value and notify the receiver. After the communication, the transactions can execute 
in active participants. 

We only described the key operations in \calvin that is relevant to 
communications and omit many details, which can be found in ~\cite{Thomson:2012:CFD:2213836.2213838}.
The main challenge of implementing \calvin is to design the 
sophisticated data structures to facilitate the correct communication 
between machines, especially for one-sided primitives. 
We choose not to discuss them in detail since it is mainly 
engineering efforts. 
Compared to other five protocols, we do not need to consider many
subtle issues to ensure correctness, because after transaction dispatch
and RS/WS preparation, the execution is mostly local. 
We believe including \calvin in \projectname is important because 
we can understand the communication implementation and cost
for the shared-nothing protocol.
As far as we know, it is also the first implementation of \calvin
with RDMA. 

\section{Hybrid Protocols}
\label{hybrid-protocol}

%\subsection{Motivation}
%Previously in this section, we have described the design and implementations of two-sided and one-side protocols.
With the ability to evaluate all protocol stages in 
\projectname,
a natural question is: what would be the best
implementation if we can use different primitives
for different stages?
 DrTM+H~\cite{wei2018deconstructing} only provides the 
 answer for \occ, but what about others?
 %has provided an answer for one OCC variant and provide some guidelines of which primitives to use for which protocol stage. However, it is still not clear what is the potential of mixing the use of primitives in different stages, especially for various protocols. 
 
 %With \projectname, users can create stage-wise latency results to pinpoint the bottleneck of a protocol and to cherry-pick the primitive that incurs a lower stage latency for any workload. 2) 
% While we have already done some experiments to compare various aspects of different protocols in \projectname, previous latency results are still not an accurate reflection of the actual bottleneck of each protocol since interleaving transactions among multiple co-routines increases transaction latency nondeterministically. 
%To pinpoint the stage-wise bottleneck of each protocol implementation further, we run experimented using {\em only one} co-routine while maintaining other experiment parameters as default. We break down a whole transaction latency into different stages: \textit{Read}, \textit{Lock}, \textit{2PC} (the \texttt{prepare} phase), \textit{Log}, \textit{Release} and \textit{Commit}. \sundial has an extra lease \textit{Renew} latency. Figure~\ref{fig:latency-breakdown} shows the stage-wise latency breakdown. Note that a record is returned together with locking result (as a  whole tuple) for \nowait and \waitdie, so their \textit{Read} latency is combined into the \textit{Lock} latency.

\begin{figure*}[ht]
    \centering
    \includegraphics[width=\linewidth]{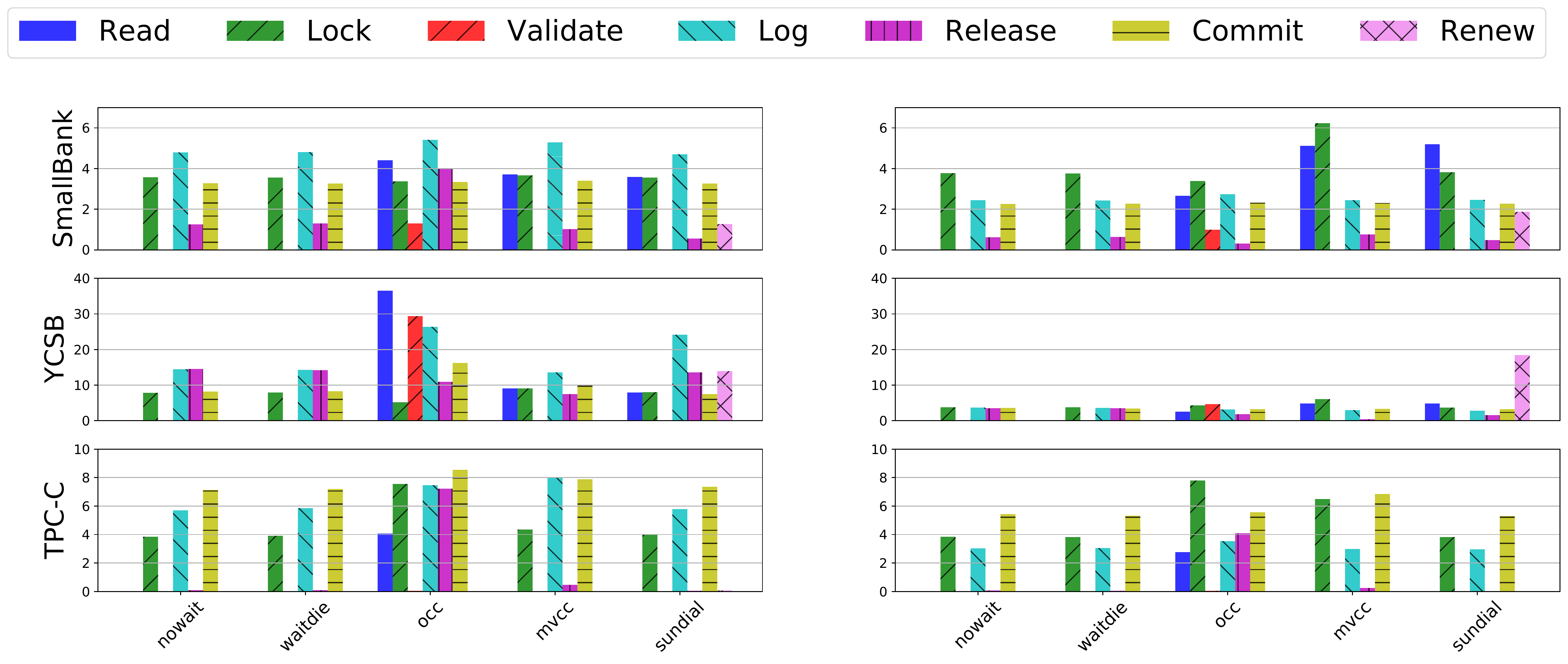}
    \vspace{-0.6cm}
    \caption{Latency breakdown: \textbf{RPC} (Left), \textbf{one-sided} (Right)}
    \vspace{-0.4cm}
    \label{fig:latency-breakdown}
\end{figure*}

\subsection{Methodology}
\projectname allows using two methods for exploiting the potential of hybrid protocols. The first method is
based on the stage-wise latency breakdown produced by \projectname. Accordingly, the hybrid designs for protocols can be straight-forward by {\em cherry-picking} the better communication type among the two-sided and one-sided world for each operation. Figure~\ref{fig:latency-breakdown} shows the latency-breakdown of all five protocols in \projectname using one co-routine under various workloads. As one example, we can see that for SmallBank: 1) a hybrid design of \mvcc which includes {RPC} Read \& Lock and {one-sided} Log \& Release \& Commit can be a good candidate of hybrid \mvcc; 2) a hybrid design of \sundial which includes {RPC} Read \& Renew and {one-sided} Lock \& Log \& Commit will incur shorter latency and thus may improve its throughput on SmallBank. 
With the analysis of latency results, we see that:
1) Log, Commit and 
Release operations prefer one-sided operations; 
2) \sundial's renew operation prefers two-sided RPC;
3) For complex read/lock operation as in \mvcc and \sundial, two-sided RPC may be rewarding; and
4) The best hybrid designs of any protocol are workload-dependent.

\begin{figure*}[ht]
    \centering
    \includegraphics[width=17cm]{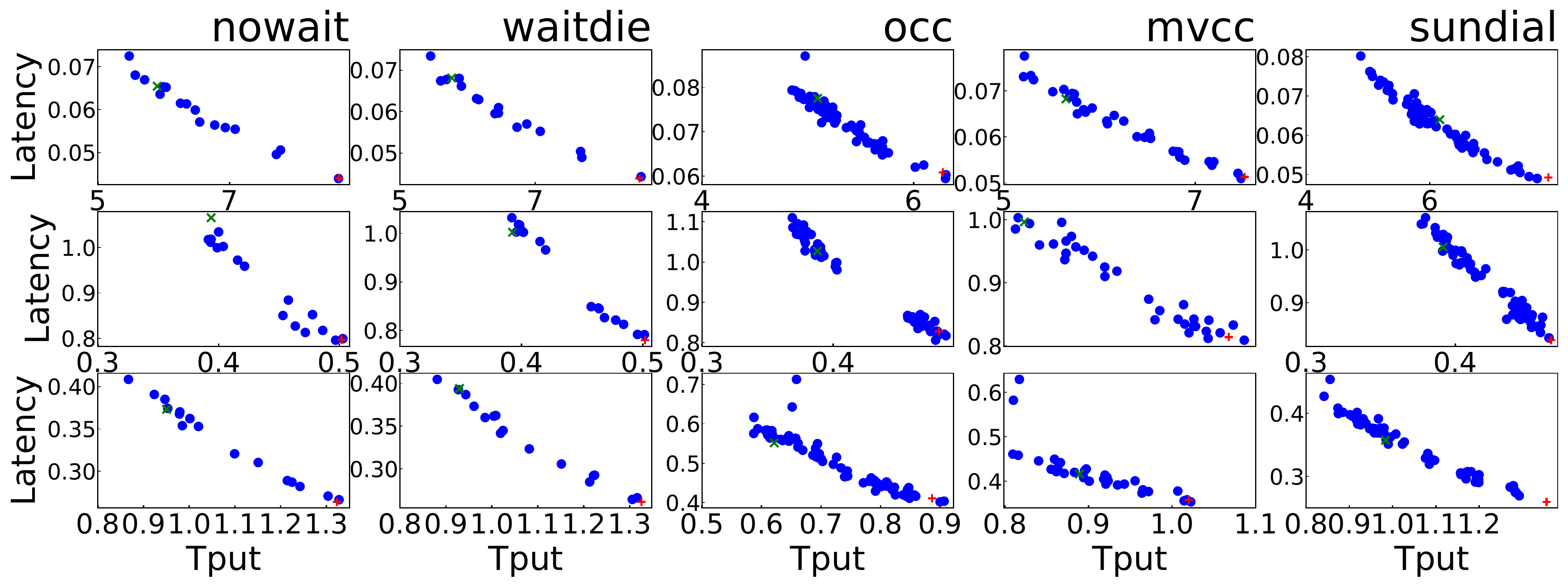}
    \vspace{-0.2cm}
    \caption{Performance of all stage-wise hybrid implementations compared with purely two-sided or purely one-sided implementation: \textbf{RPC} (Green), \textbf{one-sided} (Red), \textbf{hybrid} (Blue) for three workloads from up to down: SmallBank, YCSB, TPC-C.}
    \vspace{-0cm}
    \label{fig:hybrid_lat_tput}
\end{figure*}

Alternatively, \projectname has implemented all protocols in a way that makes it possible to 
conduct the exhaustive search of all combinations of hybrid protocols. This is useful when multiple co-routines are involved or when the system load is high. \projectname provides a configurable framework that could comprehensively evaluate \emph{any} two-sided, one-sided, and \emph{any} combination of hybrid implementations of protocols included. To implement this goal, we provide coding for each hybrid implementation, each binary digit in the code specifying the primitive to use for each stage. This interface allows \projectname to be friendly to both common and expert users: common users can find the best hybrid implementation given the protocol and the workload specification. Expert users can specify their own hybrid code to 
indicate the primitive used in each stage and verify their intuitions quickly. By leveraging \projectname, we aim to find solid evidence of the best hybrid design instead of allowing users to guess and try based on suggestive guidelines. Figure~\ref{fig:hybrid_lat_tput} shows a comparison of stage-wise hybrid protocols compared to their purely two-sided RPC or one-sided implementation when 10 co-routines are used for four protocols. It can be seen that most hybrid designs span in the middle of purely two-sided and purely one-sided designs. Yet there are hybrid ones that do outperform the better of the two both latency-wise and throughput-wise.

%\vspace{-2mm}
\subsection{Implementation Challenges}
The design of a universal hybrid implementation generator has some challenges to ensure correctness. 
First, the remote tuple address must be recorded for RPC Read or Lock. This is needed because future one-sided stages may need the offset to access the tuple. 
Second, any two-sided or one-sided stage must work correctly, assuming that it may work with another stage using a different primitive. We rely on a shared RDMA-enabled memory region for every tuple in the read/write set to maintain the correct communication between heterogeneous stages. 
Third, the heterogeneous stages must reach a consensus to indicate if one has finished its work correctly. This may cause tricky issues if not handled carefully. One example is that a lock RPC handler must notify lock requesters of the completion of the lock by not only sending back a success reply, but also writing a success bit in the agreed region in the RDMA-enable memory of the locked tuple so that one-sided Release stage can successfully release the lock.

%\subsection{Discussion}
In general, hybrid implementations may bring performance benefits in some cases. The design choice is still limited. Upon locking, one design choice for RPC-based Lock/Release is that the lock handler can CAS the requester's timestamp locally on the lock field so that the Release can just send its timestamp; a Release handler would try to actually release by CASing a zero into the lock bit if the received timestamp equals to the one record in the metadata. Failure to CAS means that the record was not locked by this requestor and was not released, which happens to be correct behavior. However, with RPC-based Lock and one-sided Release, this design choice is impossible because the Release stage must know exactly the tuples locked before issuing a one-sided WRITE for each of them.
\vspace{-2mm}
\section{Evaluation}

\begin{figure*}[t!]
    \centering
    \includegraphics[width=17cm]{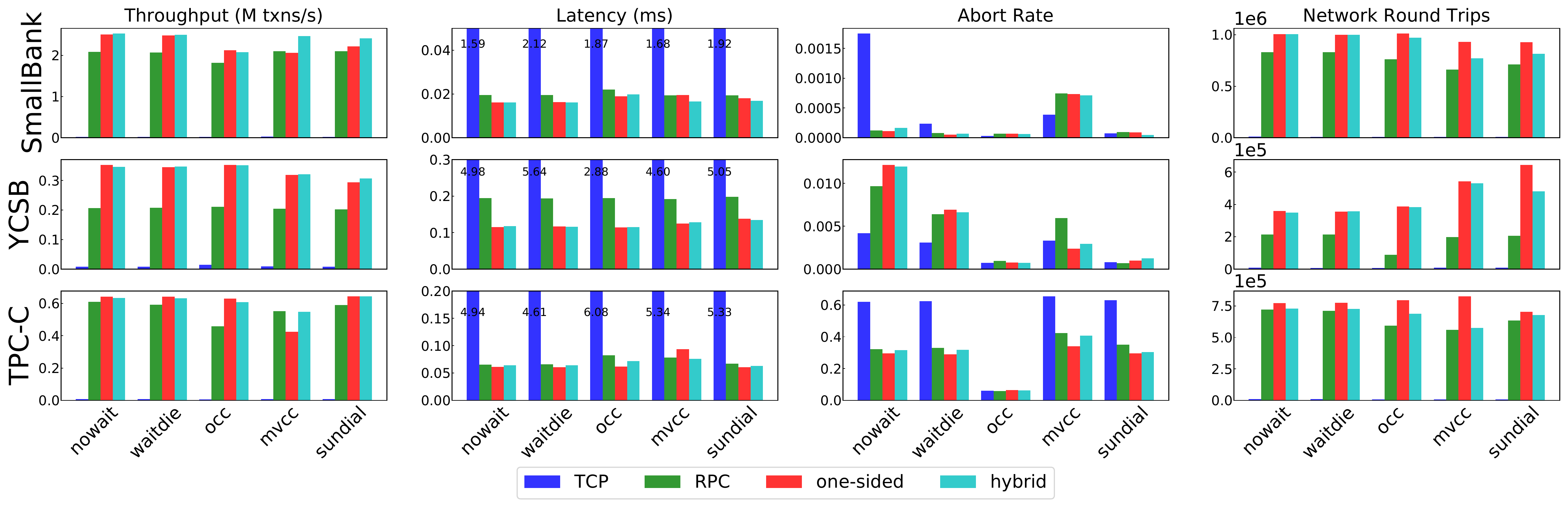}
    \vspace{-0.0cm}
    \caption{Overall throughput, latency, abort rate and \# of network round trips}
    \vspace{-0.4cm}
    \label{fig:eval_overall}
\end{figure*}

\vspace{-2mm}
\subsection{Workloads and Setups}
% \vspace{-2mm}

We use three popular OLTP benchmarks, SmallBank~\cite{SmallBank}, YCSB~\cite{cooper2010benchmarking}, and TPC-C~\cite{TPC-C}, to evaluate protocols using two-sided {\em RPC},
{\em one-sided} primitives, and 
{\em hybrid} protocol developed in Section~\ref{hybrid-protocol}.
Depending on the benchmarks, the best hybrid design choices are different. We leverage the methodology described in section~\ref{hybrid-protocol} to cherry-pick the best hybrid design choice. In some benchmark and protocol, the best hybrid design choice happens to be purely one-sided across all stages.
We include the traditional TCP-based protocols in the evaluation section. 
For all benchmarks, records are partitioned across nodes. 
%We enforce all transactions to access distributed records. %To avoid effecting the results due to the assumption of locality, 
%To eliminate the effects of locality, all transactions 
%use network operations to fetch and update the data. 

\textbf{SmallBank}~\cite{SmallBank} is a banking application. Each transaction performs reads and writes on the account data. SmallBank features a small number of writes and reads in one transaction ($<3$) with 
simple arithmetic operations, making SmallBank a network-intensive application. 
%Moreover, each record in SmallBank is of 4 bytes: it just stores simple information like checking and saving. It has six types of transactions, each sharing the same possibility of execution in our evaluations.

\textbf{YCSB}~\cite{cooper2010benchmarking} (The Yahoo! Cloud Serving Benchmark) is designed to evaluate large-scale Internet applications. There is just one table in the database. YCSB parameters such as
record size, the number of writes or reads involved in a transaction, the ratio of read/write, the contention level, and time spent at the computation phase are all configurable. In all our experiments, the record length is set to 64 bytes. The number of records in the YCSB table is proportional to the cluster size and the number of transaction threads used.
%\red{[to be deleted]The number of entries and the hot area we use is proportional to the number of threads in \projectname.} 
By default, each transaction contains 10 operations: 20\% write, and 80\% read, and it spends 5 microseconds in its execution phase. The hot area accounts for 0.1\% of total records. The contention in YCSB is controlled by allowing a configurable percentage of read/write to access the hot area, which we call the Hot Access Probability, which is 10\% probability by default. 
In Section~\ref{eval-contention}, we study the effects of different contention
levels. 
% in different experiments.
% In our test, an execution phase consuming 25\% of the whole transaction time is added to simulate the actual computation of a transaction.

\textbf{TPC-C}~\cite{TPC-C} simulates the processing of warehouse orders and is representative of CPU-intensive workloads. In our evaluation, we run the \textbf{new-order} transaction since other transactions primarily focus on local operations. The \textbf{new-order} accounts for 45\% in TPC-C and consists of longer (up to 15) distributed writes and complex transaction executions. 

%\red{In this benchmark, TPC-C has higher CPU utilization than SmallBank}.

%\vspace{-2mm}
%\subsection{Execution Setup}
%\label{eval_setup}
% \vspace{-2mm}

We evaluate \projectname on four nodes of an RDMA-capable EDR cluster, each node equipped with two 12-core Intel Xeon E5-2670 v3 processor, 128GB RAM, and one ConnectX-4 EDR 100Gb/s InfiniBand MT27700. 
%The second cluster is an RDMA-capable cluster with 16 nodes. Each node has two 8-core Intel Xeon CPU E5-2630 v3 processors, 64GB RAM, and one ConnectX-3 Pro FDR 56Gb/s InfiniBand MT27520. 
As there is only one RNIC on each node, we only run evaluations on the CPU on the same NUMA node with the RNIC to prevent NUMA from affecting our results. 
%We name the first cluster EDR and the second cluster FDR in our experiments. 
By default, we use ten transaction execution threads per node and use 1 co-routine in section~\ref{eval-overall} and 10 co-routines in section~\ref{eval-contention}, \ref{eval-computation}, and \ref{eval_qp_scalability}. 
%each co-routine producing and handling transaction requests independently. 
We enable 3-way replication for \projectname.
%: every transaction thread logs updates to two remote backup machines in the cluster before each transaction's update operation in the commit phase.
%For the FDR cluster, we use eight threads due to the limited number of cores in one CPU. 
The implementations in \projectname are evaluated on three metrics: {\em throughput, latency and abort rate}. 
%To concentrate on comparing communication behaviors and protocols, \red{we enforce coordinators or \calvin sequencers to self-generate transaction requests instead of receiving from clients}. 
%Since all protocols except \calvin are shared-everything implementations, we focus on their comparisons throughput this section. 
%For \calvin, we discuss its evaluation in Section~\ref{}.

%\vspace{-2mm}
\subsection{Overall Results}
\label{eval-overall}
% \vspace{-2mm}

Figure~\ref{fig:eval_overall} shows the
results of all three implementations of 
the six protocols. 
The results show the effects of different 
implementations and cross-protocol comparisons. 

For the same protocol, the performance
of one-side is generally better than RPC, except
MVCC under TPC-C. 
\mvcc does not benefit from one-sided primitives on TPC-C, both latency-wise and throughput-wise. As TPC-C contains long 100\% write operations, all protocols incur over 50\% abort rate. Therefore latency is determined by how quickly an abort decision can be made. one-sided \mvcc does not outperform RPC in this scenario since a one-sided \mvcc transaction may need two round trips to decide to abort.

Across all one-sided implementations, \occ is one best choice for YCSB, yet
it becomes the second-to-the-worst for SmallBank. 
In fact, {one-sided} 2PL has better performance on SmallBank over {one-sided} \occ, \mvcc and \sundial. 
% since the reduced read-write conflicts do not benefit from all-write transactions. \sundial can abort early just like \waitdie without having to wait until the commit stage for write-write conflicts.
Besides, the best protocol choice not only depends on workload characteristics but also depends on communication types. For YCSB, the performance of {RPC} implementations are similar across protocols while {one-sided} ones peaks at \occ.

As for the chosen hybrid implementations, 
%most of them perform similar to the better side of either RPC or one-sided implementations. However, 
we found three occurrences where a hybrid implementation indeed does much better than both. On SmallBank, the hybrid \mvcc performs 17.8\%  and 21.7\% better than the RPC and one-sided implementations; the hybrid \sundial performs 14.8\% and 8.6\% better than its RPC nad one-sided counterparts. On YCSB, the hybrid \sundial performs 51.6\% and 4.5\% better than the RPC and one-sided implementations.
\subsection{Effect of Co-routines}
\label{eval_coroutines}
%  \vspace{-2mm}

\begin{figure}[ht]
    \centering
    \includegraphics[width=8.4cm]{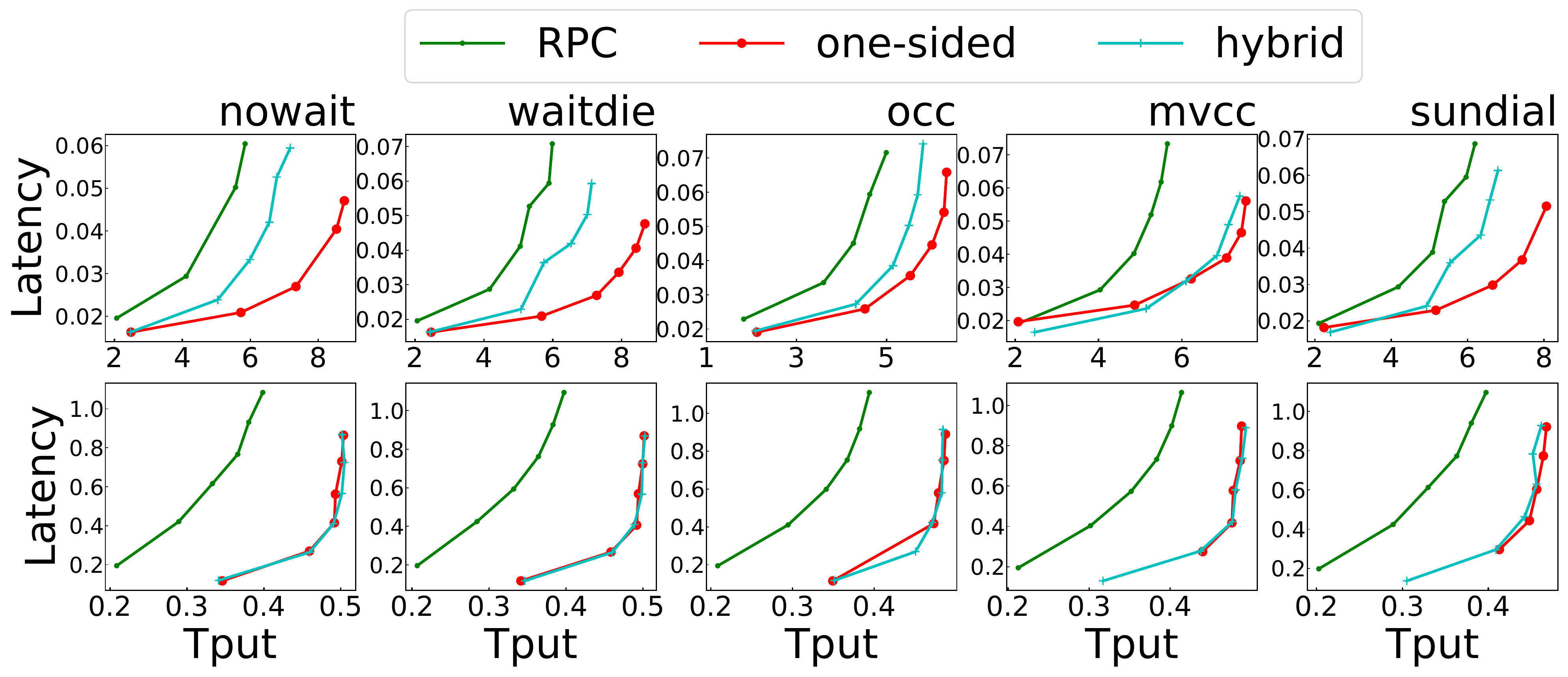}
    % \vspace{-0.2cm}
    \caption{Throughput (M txns/s) and Latency (ms) for \textbf{SmallBank} (Up) and \textbf{YCSB} (Down) with increasing co-routines.}
    \vspace{0cm}
    \label{fig:coroutine-latency-tput}
\end{figure}

% \begin{figure}[b]
%     \centering
%     \vspace{-0.4cm}
%     \includegraphics[width=\linewidth]{images/eval_coroutines.pdf}
%     \vspace{-0.6cm}
%     \caption{Effect of \#co-routine on throughput (M txns/s)}
%     \vspace{-0.2cm}
%     \label{fig:eval_cor_curve}
% \end{figure}

Figure~\ref{fig:coroutine-latency-tput} shows
the latency and throughput change when increasing 
the number of co-routines from 1 to 11 with a 
step of 2 for both SmallBank and YCSB.
We see that the latency is always increased
with more co-routines due to the 
overhead of context switches. 
Also, the throughput increases
since more co-routines can hide the latency 
of network operations. 
However, we also observe that throughput
starts to plateau after a certain number of 
co-routines. This is due to the higher 
contention with longer latency. The performance of hybrid implementations lies in the middle between RPC and one-sided ones for SmallBank and similar to the one-sided implementations for YCSB as more co-routines are used.

\begin{wrapfigure}[13]{r}{0.5\linewidth}
    \centering
    \vspace{-0.2cm}
    \includegraphics[width=\linewidth]{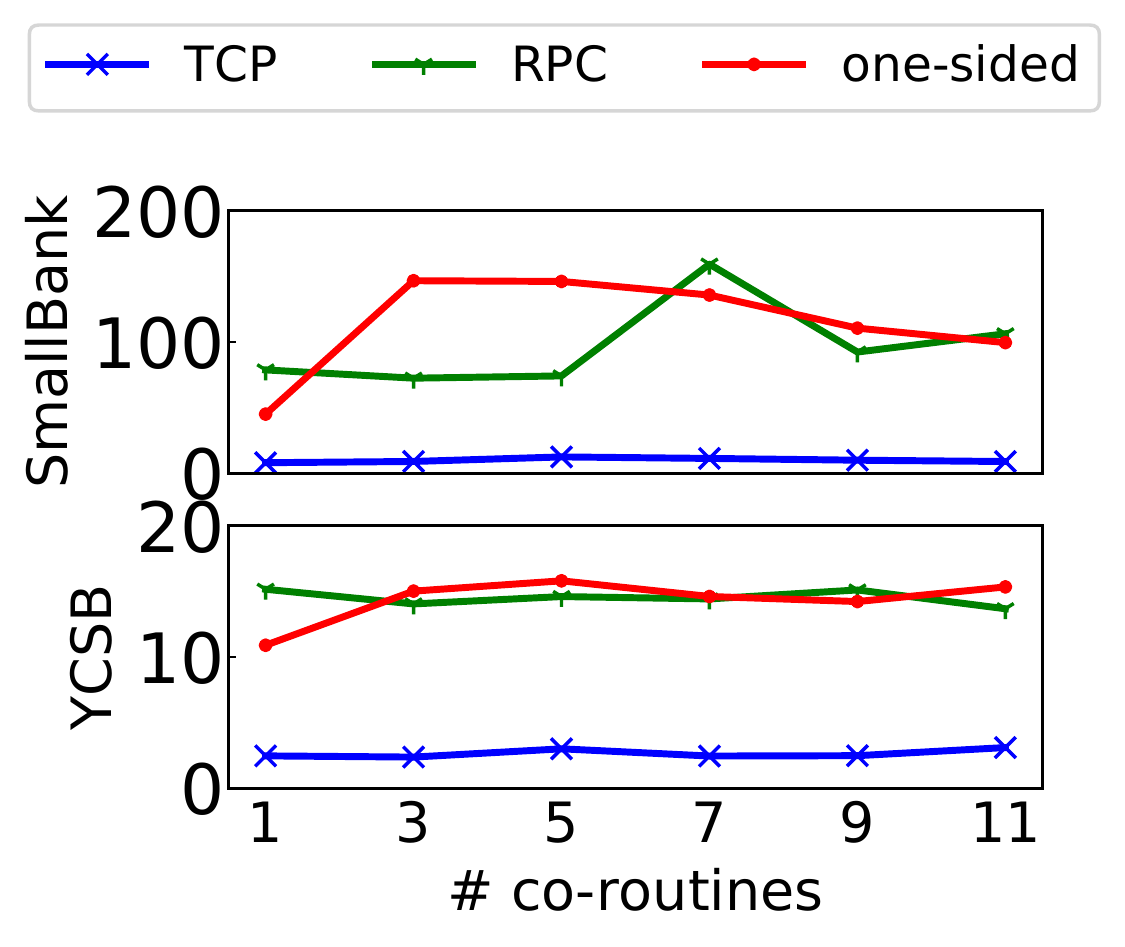}
    \vspace{-5mm}
    \caption{\calvin's throughput (K txns/s) w.r.t \#co-routines}
    \vspace{-0.1cm}
    \label{fig:eval_calvin_coroutine}
\end{wrapfigure}

Figure~\ref{fig:eval_calvin_coroutine} shows the
results for \calvin. Due to its shared-nothing
architecture, it is not directly comparable 
to others. 
In both RPC and one-sided, 
we see that increasing \#co-routines may or may not improve throughput. 
This is because \calvin requires RDMA-based epoch synchronization among all sequencer co-routines on all machines; therefore network latency due to staggered co-routines cannot be hidden with the use of increasing \#co-routines. Note that RDMA-based \calvin does not reach as higher throughput as other protocols compared to their TCP-based versions due to high synchronization cost.

 \setlength{\intextsep}{2pt}%
 \setlength{\columnsep}{8pt}%

\subsection{Effect of Contention Level}
\label{eval-contention}
%\vspace{-2mm}

\begin{figure}[htp]
    \centering
    % \vspace{-0.2cm}
    \includegraphics[width=\linewidth]{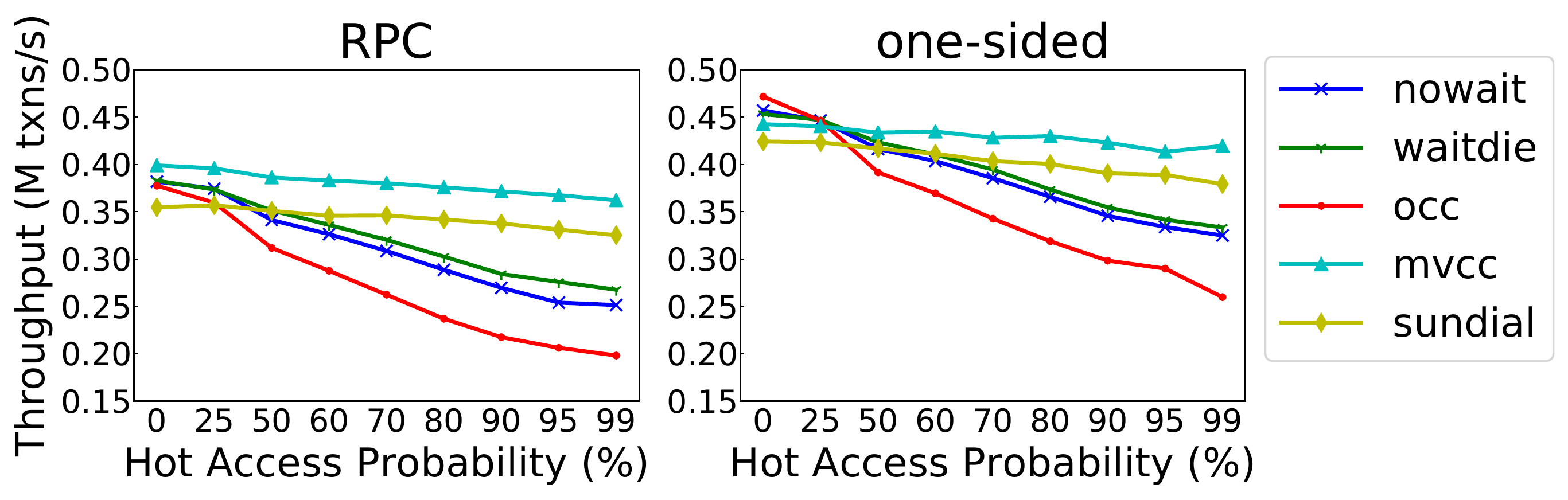}
    \vspace{-0.2cm}
    \caption{Effect of contention level on throughput for YCSB}
     \vspace{-2mm}
    \label{fig:eval_conflict_level}
\end{figure}

Figure~\ref{fig:eval_conflict_level} shows
the throughput of RPC and one-sided implementation
of different protocols with different contention 
levels using YCSB. 
%To show how well RDMA-based protocols can handle contention, we compare the throughput of protocols with different contention levels for YCSB. 
We control the contention levels by limiting the number of hot records to 0.1\% of total records and varying the possibility of one read or write visiting hot records. 
%Other YCSB configurations are as default. 
% Another parameter is the read/write ratio for one transaction. In our experiments, the read/write ratio is 1:4.
% Moreover, we also simulate execution stage by doing dummy computation which takes 5\% of the total latency of a full transaction.
%The results are shown in Figure~\ref{fig:eval_conflict_level}. 

We have several key observations. 
With low contention, the throughput differences
are small, and the worst one-sided is better
than the best RPC. 
As the contention increases, the thoughput
of all protocols decrease, but \occ always
drops most significantly because of a larger possibility to abort and high abort cost due to its optimistic assumption under a high contention level. 
The performance of \nowait and \waitdie 
also decrease considerably due to the intensive conflict read and write locks. \mvcc and \sundial are less affected when the conflict rate increases.
As a result, with high contention, 
the throughput of different protocols
become quite different, but the gaps between 
RPC and one-sided are much smaller.
We also notice that one-sided \sundial and \mvcc, although featuring advanced read-write conflict management, are worse than one-sided \occ 
at low contention. It is because these two have more complicated operations to maintain more information to reduce the abort rate, which is more costly. After all, every access to remote data will trigger network operation in their one-sided versions.
A key conclusion is that \occ is not the best---in fact always the worst with high contention,
it justifies our study of different protocols
with a common framework.

\subsection{Effect of Computation}
\label{eval-computation}
%\vspace{-2mm}

\begin{figure}[htp]
    \centering
    % \vspace{-0.2cm}
    \includegraphics[width=\linewidth]{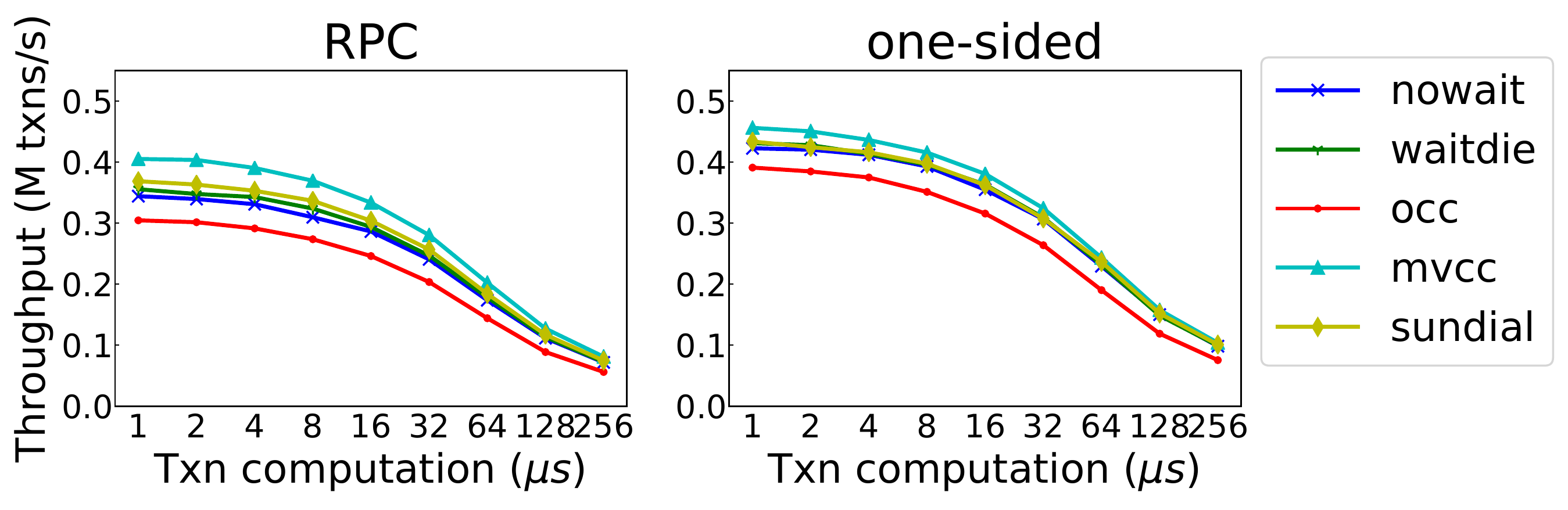}
    \vspace{-0.5cm}
    \caption{Effect of computation on throughput for YCSB}
    \vspace{-0.0cm}
    \label{fig:eval_sleep_curve}
\end{figure}

% As shown in the overall results, \textbf{one-sided} version can better handle transactions bounded by computation. 
%So we increase the percentage of the computation in one transaction. 
To study the effect of different computation time in the whole life of transaction execution, we add dummy computation in the execution stage of YCSB, ranging from 1 to 256 $\mu$s. We show results in Figure~\ref{fig:eval_sleep_curve}. We observe that 
(1) {RPC} and {one-sided} share a similar decreasing trend as computation increases; and 
(2) the advantage of one-sided over RPC is diminishing as the 
computation workloads increase.
For RPC, more computation will increase the latency to handle RPC 
request; for one-sided, more computation will narrow its advantage 
over RPC due to the non-involvement of CPU in communication.

\begin{figure}[ht]
    \centering
    \includegraphics[width=\linewidth]{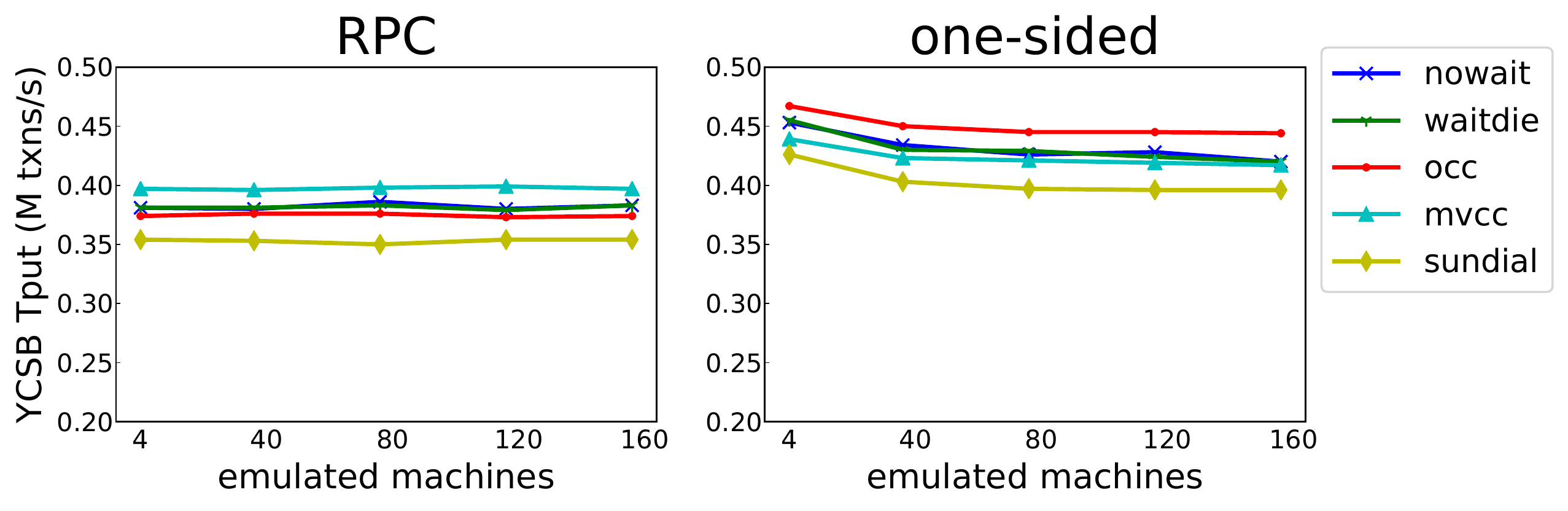}
    \vspace{-0.6cm}
    \caption{Throughput on emulated large EDR clusters.}
    %\vspace{-0.2cm}
    \label{fig:qps-scalability}
\end{figure}

%\vspace{-2mm}
\subsection{Scalability of QPs}
\label{eval_qp_scalability}
%\vspace{-2mm}

To understand how protocols perform with a much larger cluster, we run all protocols against the YCSB benchmark
(90\% hot access probability and 0.1\% hot area) 
on several emulated larger EDR clusters.
%each equipped with enough QPs for $4x$ number of fully connected nodes where $x=1,10,20,30,40$,
as shown in Figure~\ref{fig:qps-scalability}. 
%Note that UD QPs are per-node for two-sided RPC, while RC QPs are per-node and per-thread for one-sided primitives. 
Each RDMA op selects the sending QP from multiple same-destination QPs in a round-robin manner to emulate the network traffic on large clusters.
We observe that on emulated larger clusters, one-sided implementations maintain their
superiority over RPC ones up to a 160-node cluster, yet the advantage gap closes as cluster size increases. We attribute this behavior to the fact that an increasing number of QPs needed for larger clusters will cause performance loss due to limited NIC capabilities. 

\vspace{-2mm}
\section{Related Work}
% \vspace{-2mm}
{\bf Comparisons among concurrency control protocols}
%Prior work has made valuable comparisons on different concurrency control protocols. 
\cite{agrawal1987concurrency} uses modeling techniques to reveal the hidden connections between protocols' underlying assumptions as well as their seemingly contradictory performance results.
\cite{huang1991experimental} compares three concurrency control protocols in real-time database systems but only restraints to optimistic ones. \cite{yu2015evaluation,yu2014staring} focuses on the scalability issues and examines seven concurrency control protocols on a main-memory DBMS on top of a simulated 1024-core system.  Deneva~\cite{harding2017evaluation} is the recent work comparing distributed concurrency control protocols in a single unified framework. 
%Deneva executes all transactions as stored procedures that run on servers, and analyzes the performance of different protocols under different settings like network speed, node number, update rate and contention. 
\projectname 
takes the first step in comparing different protocols under the context of various RDMA primitives.

{\bf Comparisons between RDMA primitives}
%Before \projectname, there has already been work comparing RDMA primitives. 
\cite{kaminsky2014using} compares the use of RDMA \texttt{WRITE} and RDMA \texttt{READ} when constructing a high performance key-value system. \cite{dragojevic2014farm} finds out that RDMA \texttt{WRITE}'s polling significantly outperforms \texttt{SEND} and \texttt{RECV} verbs when constructing the FaRM's communication subsystem. \cite{kalia2016fasst} shows that UD-based RPC using \texttt{SEND} and \texttt{RECV} outperforms one-sided primitives. \cite{wei2018deconstructing} did more primitive-level comparisons with different payload sizes. Compared
to them, \projectname compares the primitives
with a much wider range of concurrency control algorithms
on two clusters with different RDMA capabilities.

%applicRDMA nation level in order to reach useful insights in constructing concurrency control protocols.

{\bf Distributed transaction systems}
High performance transaction systems have been investigated intensively~\cite{Thomson:2012:CFD:2213836.2213838,corbett2013spanner,tu2013speedy,dragojevic2015no,chen2016fast,wei2018deconstructing,lee2015implementing}. Most of them focus on distributed transaction systems~\cite{corbett2013spanner,dragojevic2015no,chen2016fast,wei2018deconstructing} since it is more challenging to implement a high performance transaction system with data partitioned across the nodes. 
Some works, e.g., \cite{lee2015implementing,dragojevic2015no,chen2016fast,wei2018deconstructing,kalia2016fasst}, focus only on one protocol (i.e., some variants of \occ). Other works like~\cite{wei2015fast,yu2018sundial,Thomson:2012:CFD:2213836.2213838} explore novel techniques like determinism or leasing. However, these works did not explore the opportunity of using RDMA networks.

{\bf RDMA-based database systems}
NAM-DB\cite{zamanian2016end} is a scalable database system that employs RDMA. First, it leverages one-sided primitives to implement the snapshot isolation (SI) protocol. Compared to NAM-DB, \projectname focuses on protocols that guarantee serializability, which is a more strict isolation level that frees users from the burden of maintaining non-trivial constraints. Second, NAM-DB\cite{zamanian2016end} employs scalable timestamp generation to optimize SI, while our timestamp is generated by using the local clock while allowing protocols to handle the skew when needed.

{\bf Deterministic Concurrency Control}
Determinism is a popular idea in the community of concurrency control. 
% Calvin\cite{Thomson:2012:CFD:2213836.2213838} leverages one scheduler thread per machine to enforce a deterministic serializable order on a batch of transactions in one epoch, thus avoiding non-deterministic aborts. However, as a distributed concurrency control protocol, Calvin is an inherent shared-nothing design, and therefore the transaction determinism across machines comes at the cost of sequencing and synchronization overhead per batch. On the contrary, protocols in \projectname are all implemented as shared-everything designs. Therefore, Calvin is naturally suitable for \cite{harding2017evaluation} as one comparison candidate, while it is not fair to be directed compared within \projectname.
Bohm\cite{faleiro2014rethinking} implements serializable multi-version concurrency control. Unlike \calvin implemented in \projectname, Bohm introduces determinism into the multi-core environment, assigning records to deterministic threads in the concurrency control layer to avoid aborts. 
% \projectname 
% has not included Bohm because
% \projectname right now focuses on distributed concurrency control protocols.
%scalability concerns arise when implementing Bohm in the distributed setting: mapping all distributed records deterministically to concurrency control threads in one node will not scale with an increasing number of partitions; Mapping all records deterministically to inter-node threads may incur high synchronization overhead among nodes. Bohm threads synchronize by using a barrier within a multi-core environment and thus is much simpler. 
We leave the efficient implementation of Bohm in the distributed setting using RDMA as an important future work for \projectname.

%We didn't include these protocols in \projectname because they didn't naively line up with \projectname's current distributed execution model. Calvin\cite{} requires the use of an extra sequencer thread and scheduler thread, which can potentially become the bottleneck of the system. while 
%Bohm\cite{} is not trivial to be implemented in a distributed setting. }
%Our preliminary results show that Calvin cannot benefit much from RDMA because of the use of sequencer thread and scheduler thread as well as frequent broadcasting among participants.}
%\vspace{-2mm}
\section{Conclusion}
%\vspace{-2mm}

We develop
\projectname, the first
unified and comprehensive RDMA-enabled distributed transaction processing framework supporting six serializable concurrency control protocols---not only the classical protocols \nowait, \waitdie and \occ, but also more advanced \mvcc and \sundial, and 
even \calvin---the deterministic concurrency control protocol.
Our goal is to unbiasedly compare the protocols 
in a common execution environment with the concurrency 
control protocol being the only changeable component. 
Based on \projectname, we get the deep insights
that cannot be obtained by any existing systems.
Most importantly, we obtain the execution stage latency 
breakdowns with one-sided and two-sided primitive 
for each protocol, which are
analyzed to develop the efficient 
hybrid implementations. Moreover, \projectname can enumerate all hybrid implementations for protocols included under given workload characteristic.
%---for each
%stage we should choose the primitives that perform better.
%We prototype the hybrid implementation
%for each protocol, which indeed achieves better performance.
\projectname is a significant advance
over the state-of-the-art; it can both provide performance insights
and be used as the common infrastructure for fast
prototyping new two-sided, one-sided, and hybrid implementations. 

\bibliographystyle{ACM-Reference-Format}
\bibliography{main}

%%
%% If your work has an appendix, this is the place to put it.
% \appendix

% \section{Research Methods}

% \subsection{Part One}

% Lorem ipsum dolor sit amet, consectetur adipiscing elit. Morbi
% malesuada, quam in pulvinar varius, metus nunc fermentum urna, id
% sollicitudin purus odio sit amet enim. Aliquam ullamcorper eu ipsum
% vel mollis. Curabitur quis dictum nisl. Phasellus vel semper risus, et
% lacinia dolor. Integer ultricies commodo sem nec semper.

% \subsection{Part Two}

% Etiam commodo feugiat nisl pulvinar pellentesque. Etiam auctor sodales
% ligula, non varius nibh pulvinar semper. Suspendisse nec lectus non
% ipsum convallis congue hendrerit vitae sapien. Donec at laoreet
% eros. Vivamus non purus placerat, scelerisque diam eu, cursus
% ante. Etiam aliquam tortor auctor efficitur mattis.

% \section{Online Resources}

% Nam id fermentum dui. Suspendisse sagittis tortor a nulla mollis, in
% pulvinar ex pretium. Sed interdum orci quis metus euismod, et sagittis
% enim maximus. Vestibulum gravida massa ut felis suscipit
% congue. Quisque mattis elit a risus ultrices commodo venenatis eget
% dui. Etiam sagittis eleifend elementum.

% Nam interdum magna at lectus dignissim, ac dignissim lorem
% rhoncus. Maecenas eu arcu ac neque placerat aliquam. Nunc pulvinar
% massa et mattis lacinia.

\end{document}